\documentclass{article}

\usepackage{arxiv}
\usepackage{natbib}
\usepackage[utf8]{inputenc} % allow utf-8 input
\usepackage[T1]{fontenc}    % use 8-bit T1 fonts
\usepackage{url}            % simple URL typesetting
\usepackage{booktabs}       % professional-quality tables
\usepackage{amsfonts}       % blackboard math symbols
\usepackage{nicefrac}       % compact symbols for 1/2, etc.
\usepackage{microtype}      % microtypography
\usepackage{lipsum}
\usepackage{graphicx}
\usepackage{amsmath}
\usepackage{float}
\usepackage{mathtools}
\usepackage{amssymb}
\usepackage{algorithm}
\usepackage{algpseudocode}
\usepackage{amsfonts}
\usepackage{array}
\usepackage{booktabs}
\usepackage{arydshln}

\usepackage{multirow}

\usepackage{tabularx}

\title{Deep material network for homogenization of piezoelectric composites}

\author{
 Ting-Ju Wei \\
  Department of Civil Engineering\\
  National Taiwan University\\
  Taipei, Taiwan \\
   \And
 Yen-Ming Lu \\
  Department of Civil Engineering\\
  National Taiwan University\\
  Taipei, Taiwan \\
   \And
 Chuin-Shan Chen\thanks{Corresponding author. Email: \texttt{dchen@ntu.edu.tw}} \\
  Department of Civil Engineering\\
  Department of Materials Science and Engineering\\
  National Taiwan University\\
  Taipei, Taiwan \\
}

\begin{document}
\maketitle
\begin{abstract}

Piezoelectric composites are widely used in sensors, actuators, transducers, and energy-harvesting devices because their effective electromechanical performance can be tailored by combining constituent phases and microstructural architecture. However, conventional computational homogenization based on direct numerical simulation (DNS) is computationally expensive, particularly for multiscale simulations and material design tasks that require repeated homogenization analyses. To address this limitation, this work proposes a piezoelectric deep material network (PDMN) to efficiently homogenize two-phase piezoelectric composites. The proposed framework embeds the governing electromechanical homogenization relations directly into the network architecture, yielding a physics-informed, semi-analytical surrogate that explicitly captures the two-way coupling between the mechanical and electrical fields across constituent phases. The network is trained offline on linear electroelastic datasets and, through a fully coupled Newton--Raphson solution with a consistent electromechanical tangent, subsequently used for efficient online prediction under broader constitutive settings, including nonlinear electroelasticity and history-dependent responses. The framework is validated on two-phase composites of polyvinylidene fluoride (PVDF) and lithium niobate (LiNbO$_3$) with reversed phase arrangements under nonlinear electroelastic loading, and on a viscoelastic--piezoelectric composite exhibiting coupled stress relaxation. Numerical examples show that the proposed PDMN achieves high predictive accuracy while reducing the computational cost by more than three orders of magnitude compared with DNS. The proposed framework, therefore, provides an efficient and reliable surrogate for the multiscale analysis and design of piezoelectric composites.

\end{abstract}

% keywords can be removed

\section{Introduction}

Piezoelectric composites interconvert mechanical and electrical energy and are used in sensors, actuators, transducers, and energy-harvesting systems~\cite{sony2019literature, ferreira2022embedded, mahapatra2021piezoelectric, arnau2004piezoelectric}. By combining constituent phases with distinct electromechanical roles, such composites reach effective properties that are difficult to obtain with monolithic piezoelectric ceramics~\cite{shieh2007switching,shieh2010influence} or polymers alone. The phase interactions and microstructural morphology can be tuned for application-specific combinations of stiffness, dielectric response, and piezoelectric performance~\cite{adeniyi2021multi, abedi2020effective, habib2022review, heywang2008piezoelectricity}.

To predict the effective behavior of piezoelectric composites, computational homogenization constructs a representative volume element (RVE) of the heterogeneous microstructure. A variety of micromechanical and numerical approaches have been developed for this purpose~\cite{noorizadegan2022piezo}. On the one hand, direct numerical simulation (DNS) based on the finite element method (FEM) accurately resolves the local electromechanical fields in complex microstructures~\cite{adeniyi2021multi, yazdanparast2023determining, koutsawa2018overall, torquato2021nonlocal}. However, such full-field simulations quickly become computationally prohibitive, especially when repeated RVE analyses are required for material design, uncertainty quantification, or concurrent multiscale simulations. On the other hand, mean-field schemes such as the Mori--Tanaka method and self-consistent approaches cost far less and are widely used for composite materials~\cite{li1998micromechanics, wu2000closed, wang2022extended, chatzigeorgiou2019micromechanical, martinez2017homogenization}. Nevertheless, their applicability may be limited when the microstructural geometry is highly complex, the phase distribution is strongly heterogeneous, or nonlinear constitutive behavior becomes important.

To bridge the gap between computational efficiency and predictive capability, deep material networks (DMNs) are a surrogate modeling framework for multiscale homogenization~\cite{su2022multiscale} and have been effective for mechanical composites, particularly short-fiber-reinforced materials~\cite{liu2019deep,liu2019exploring,gajek2020micromechanics,noels2022micromechanics,wei2025foundation,jean2024graph}. The central idea of DMN is to embed homogenization theory directly into the network architecture, resulting in a physics-informed and semi-analytical model that preserves the essential micromechanical interactions within heterogeneous materials~\cite{wei2026deep}. The offline training stage uses only linear-elastic data, and the trained network is then deployed to efficiently predict nonlinear behavior online~\cite {srinivas2026rapid,wan2024decoding}. DMNs have since been extended to other material systems, including polycrystalline materials~\cite{wei2025orientation,wei2025foundation2,wei2025crystallographic}, woven composites~\cite{shin2024deep2}, porous materials~\cite{noels2022interaction}, multiphysics problems involving coupled thermomechanical responses~\cite{shin2024deep,li2024micromechanics}, and uncertainty quantification~\cite{robertson2025microstructure,wu2025stochastic}. They have also been applied to multiscale simulations of industrial components~\cite{gajek2021fe,gajek2021efficient,wei2023ls}.

Despite these advances, existing DMN formulations have been developed almost exclusively for purely mechanical material systems. The multiphysics extensions reported to date address either uncoupled transport problems, such as thermal conduction, or thermomechanical behavior, in which the coupling is introduced through an eigenstrain-type thermal-expansion contribution~\cite{shin2024deep2,shin2024deep,gajek2022fe}. Piezoelectricity is fundamentally different: the mechanical and electrical fields are bidirectionally coupled through the piezoelectric tensor, so that the stress depends on the electric field and the electric displacement on the strain, both mediated by the same coupling operator. As a result, the generalized constitutive operator is non-symmetric and indefinite, and neither the building-block solution nor the homogenization solver of the mechanical DMN, both of which rely on a symmetric positive-definite stiffness, carries over directly. To the best of our knowledge, a DMN that consistently embeds this two-way electromechanical coupling has not yet been reported.

In this work, we develop a piezoelectric deep material network (PDMN) to homogenize two-phase piezoelectric composites. Building on the interaction-based DMN formulation~\cite{gajek2020micromechanics,noels2022micromechanics}, in which interfacial interaction variables enforce the Hill--Mandel condition, the main contributions of this work are threefold. First, we derive an analytical binary homogenization block for the fully coupled stress--charge response, in which the generalized constitutive matrix assembled from the elastic, piezoelectric, and dielectric moduli is non-symmetric, and we embed this block recursively to obtain a physics-informed, semi-analytical surrogate whose parameters retain a clear microstructural interpretation. Second, we solve the coupled nonlinear localization problem online through a fully coupled Newton--Raphson scheme and derive the corresponding consistent electromechanical tangent, including its cross-coupling blocks, as required for two-scale finite-element (FE$^2$) analyses. Third, we show that a network trained only on linear electroelastic stiffness data extrapolates accurately to nonlinear electroelasticity and to history-dependent behavior, the latter demonstrated on a viscoelastic--piezoelectric composite. Numerical results show that the proposed model achieves more than three orders of magnitude speed-up over DNS while maintaining high predictive accuracy for the effective composite response, providing an efficient and reliable tool for the multiscale analysis and design of piezoelectric composites.

\section{Piezoelectric deep material network}
\label{sec02}

\begin{figure}[htbp]
\centering
\includegraphics[width=0.9\linewidth]{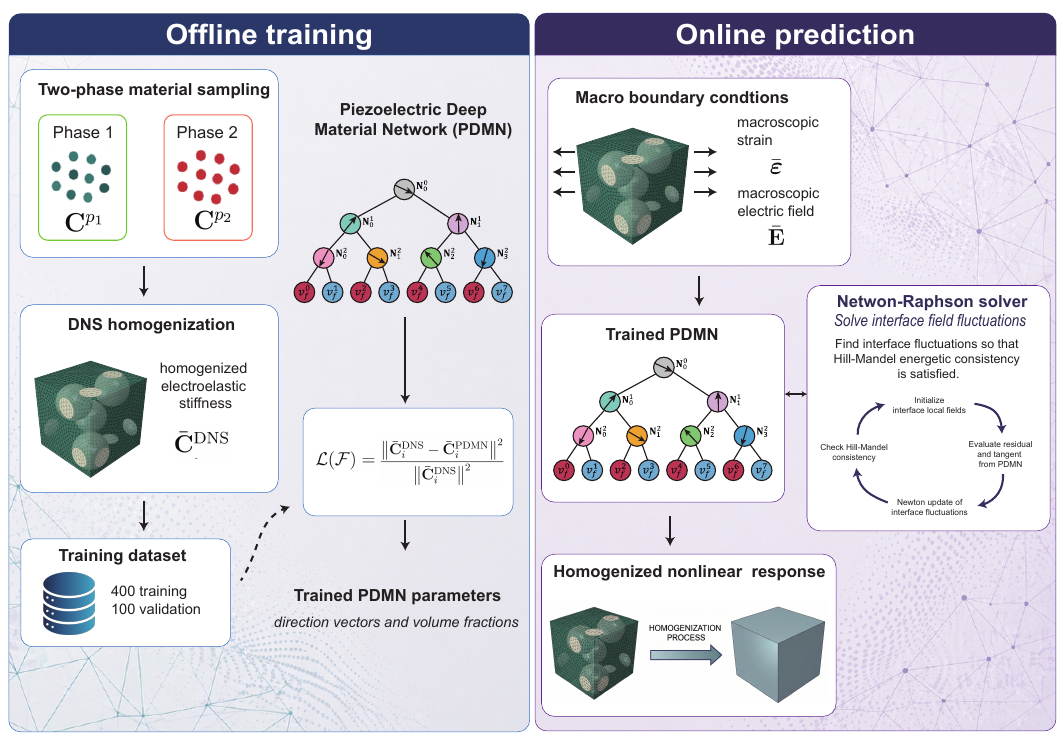}
\caption{Overview of the offline training and online prediction procedures of the proposed PDMN.}
\label{fig:Fig1_Overall}
\end{figure}

This section presents the formulation of the PDMN. We begin with the averaging theorems and the Hill--Mandel principle, which provide the theoretical foundation for representing the electromechanical response of an RVE through a set of non-overlapping subdomains. The subdomain decomposition is then restricted to a hierarchical binary partition, from which a binary homogenization building block is derived, and recursively extending this building block throughout the hierarchy yields the proposed PDMN. The subsequent subsections describe the network parameterization, field localization relations, interaction equilibrium conditions, the coupled Newton--Raphson scheme, and the consistent tangent operators required for nonlinear online constitutive updates. Finally, the offline training and online prediction procedures are introduced to complete the computational framework for evaluating the nonlinear electromechanical response of the RVE. A schematic overview of the formulation and its computational workflow is provided in Fig.~\ref{fig:Fig1_Overall}.

\subsection{Homogenization theory for reduced microstructures}

\begin{figure}[htbp]
\centering
\includegraphics[width=\linewidth]{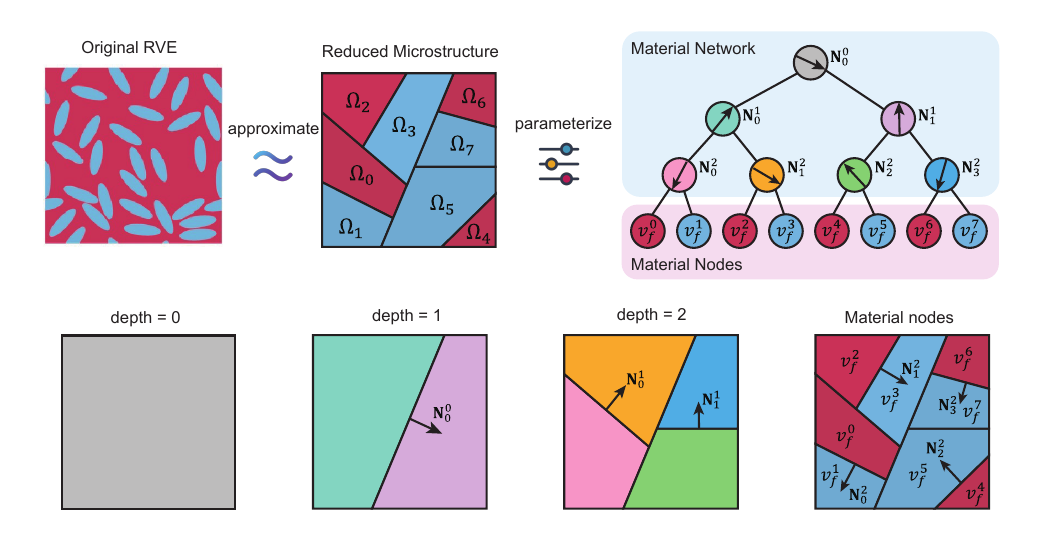}
\caption{
Schematic of the PDMN framework: the original RVE is approximated by a reduced
microstructure whose geometry is parameterized by the PDMN.
}
\label{fig:Fig2_reducedMicrostructure}
\end{figure}

Consider an RVE defined over the domain $\Omega \subset \mathbb{R}^3$.
To approximate its homogenized response, the domain is partitioned into $N_p$ non-overlapping subdomains $\Omega_\alpha$, such that the entire domain is represented by their union, as illustrated schematically in
Fig.~\ref{fig:Fig2_reducedMicrostructure}:

\begin{equation}
    \Omega = \bigcup_{\alpha=0}^{N_p-1}\Omega_\alpha,
    \qquad
    \Omega_\alpha \cap \Omega_\beta = \emptyset
    \quad \text{for } \alpha \neq \beta.
\end{equation}

According to the fundamental averaging theorems of homogenization, the macroscopic strain tensor $\bar{\boldsymbol{\varepsilon}}$ and macroscopic stress tensor $\bar{\boldsymbol{\sigma}}$
are defined as the volume averages of the corresponding local fields:

\begin{equation}
\label{eq:averaging_theorems}
\bar{\boldsymbol{\varepsilon}}
= \langle \boldsymbol{\varepsilon}(\mathbf{X}) \rangle
= \sum_{\alpha=0}^{N_p-1} v_f^\alpha \boldsymbol{\varepsilon}^{\alpha},
\qquad
\bar{\boldsymbol{\sigma}}
= \langle \boldsymbol{\sigma}(\mathbf{X}) \rangle
= \sum_{\alpha=0}^{N_p-1} v_f^\alpha \boldsymbol{\sigma}^{\alpha},
\end{equation}
where $v_f^\alpha = V_\alpha / V_\Omega$ denotes the volume fraction of
subdomain $\Omega_\alpha$, with $V_\alpha$ and $V_\Omega$ the volumes of $\Omega_\alpha$ and the entire RVE domain $\Omega$, respectively.

The energetics of the mechanical homogenization framework are governed by the
Hill--Mandel condition, which requires the macroscopic virtual work density to
equal the volume average of the microscopic virtual work density:
\begin{equation}
\label{eq:mech_hill_mandel}
\bar{\boldsymbol{\sigma}} : \delta \bar{\boldsymbol{\varepsilon}}
=
\left\langle
\boldsymbol{\sigma}(\mathbf{X})
:
\delta \boldsymbol{\varepsilon}(\mathbf{X})
\right\rangle
=
\sum_{\alpha=0}^{N_p-1}
v_f^\alpha
\left(
\boldsymbol{\sigma}^{\alpha}
:
\delta \boldsymbol{\varepsilon}^{\alpha}
\right).
\end{equation}

To map the macroscopic kinematics to the microscale, the displacement field
$\mathbf{u}(\mathbf{X})$ at any point $\mathbf{X}\in\Omega$ is decomposed into
a macroscopic affine part and a local fluctuation field:
\begin{equation}
\label{eq:displacement_decomposition}
\mathbf{u}(\mathbf{X})
=
\bar{\boldsymbol{\varepsilon}} \cdot (\mathbf{X} - \mathbf{X}_{\mathrm{ref}})
+
\boldsymbol{\xi}(\mathbf{X}),
\end{equation}
where $\mathbf{X}_{\mathrm{ref}}$ denotes a reference position vector and
$\boldsymbol{\xi}(\mathbf{X})$ represents the displacement fluctuation field
arising from microstructural heterogeneity.

The local infinitesimal strain tensor field
$\boldsymbol{\varepsilon}(\mathbf{X})$ is defined as the symmetric gradient of
the displacement field. Using the symmetric tensor product operator
$\otimes^s$, defined for two vectors $\mathbf{A}$ and $\mathbf{B}$ as
\begin{equation}
\mathbf{A}\otimes^s\mathbf{B}
=
\frac{1}{2}
\left(
\mathbf{A}\otimes\mathbf{B}
+
\mathbf{B}\otimes\mathbf{A}
\right),
\end{equation}
The strain field is obtained by substituting
Eq.~\eqref{eq:displacement_decomposition}:
\begin{equation}
\label{eq:strain_definition}
\boldsymbol{\varepsilon}(\mathbf{X})
=
\nabla\otimes^s\mathbf{u}(\mathbf{X})
=
\bar{\boldsymbol{\varepsilon}}
+
\nabla\otimes^s\boldsymbol{\xi}(\mathbf{X}).
\end{equation}

Applying the averaging operator $\langle\cdot\rangle$ to
Eq.~\eqref{eq:strain_definition} over the partitioned subdomains gives
\begin{align}
\bar{\boldsymbol{\varepsilon}}
&=
\frac{1}{V_\Omega}
\sum_{\alpha=0}^{N_p-1}
\int_{\Omega_\alpha}
\left(
\bar{\boldsymbol{\varepsilon}}
+
\nabla\otimes^s\boldsymbol{\xi}(\mathbf{X})
\right)
\,d\Omega_\alpha
\nonumber\\
&=
\sum_{\alpha=0}^{N_p-1}
v_f^\alpha
\bar{\boldsymbol{\varepsilon}}
+
\sum_{\alpha=0}^{N_p-1}
v_f^\alpha
\left[
\frac{1}{V_\alpha}
\int_{\Omega_\alpha}
\nabla\otimes^s\boldsymbol{\xi}(\mathbf{X})
\,d\Omega_\alpha
\right]
\nonumber\\
&=
\sum_{\alpha=0}^{N_p-1}
v_f^\alpha
\underbrace{
\left[
\bar{\boldsymbol{\varepsilon}}
+
\frac{1}{V_\alpha}
\int_{\Omega_\alpha}
\nabla\otimes^s\boldsymbol{\xi}(\mathbf{X})
\,d\Omega_\alpha
\right]
}_{\boldsymbol{\varepsilon}^{\alpha}}.
\end{align}

By comparison with the averaging relation in
Eq.~\eqref{eq:averaging_theorems}, the average strain in the $\alpha$-th subdomain is obtained as

\begin{equation}
\label{eq:subdomain_strain}
\boldsymbol{\varepsilon}^{\alpha}
=
\bar{\boldsymbol{\varepsilon}}
+
\frac{1}{V_\alpha}
\int_{\Omega_\alpha}
\nabla \otimes^s \boldsymbol{\xi}(\mathbf{X}) \, d\Omega_\alpha .
\end{equation}

For the electrical response, the macroscopic electric field
$\bar{\mathbf{E}}$ and the macroscopic electric displacement
$\bar{\mathbf{D}}$ are defined as the volume averages of the corresponding local fields:

\begin{equation}
\label{eq:elec_averaging_theorems}
\bar{\mathbf{E}}
=
\langle \mathbf{E}(\mathbf{X}) \rangle
=
\sum_{\alpha=0}^{N_p-1}
v_f^\alpha
\mathbf{E}^{\alpha},
\qquad
\bar{\mathbf{D}}
=
\langle \mathbf{D}(\mathbf{X}) \rangle
=
\sum_{\alpha=0}^{N_p-1}
v_f^\alpha
\mathbf{D}^{\alpha}.
\end{equation}

The electrical homogenization is governed by the corresponding Hill--Mandel condition, which states that the macroscopic electrical virtual work density equals the volume average of its microscopic counterpart:

\begin{equation}
\label{eq:elec_hill_mandel}
\bar{\mathbf{D}} \cdot \delta \bar{\mathbf{E}}
=
\left\langle
\mathbf{D}(\mathbf{X}) \cdot
\delta \mathbf{E}(\mathbf{X})
\right\rangle
=
\sum_{\alpha=0}^{N_p-1}
v_f^\alpha
\left(
\mathbf{D}^{\alpha} \cdot
\delta \mathbf{E}^{\alpha}
\right).
\end{equation}

Analogous to the displacement field, the electric potential $\phi(\mathbf{X})$ is decomposed into a macroscopic linear component and a local fluctuation field $\tilde{\phi}(\mathbf{X})$:

\begin{equation}
\label{eq:potential_decomp}
\phi(\mathbf{X})
=
-\bar{\mathbf{E}}
\cdot
(\mathbf{X}-\mathbf{X}_{\mathrm{ref}})
+
\tilde{\phi}(\mathbf{X}).
\end{equation}

The local electric field is defined as the negative gradient of the potential:
\begin{equation}
\mathbf{E}(\mathbf{X}) = -\nabla \phi(\mathbf{X}).
\end{equation}

Substituting Eq.~\eqref{eq:potential_decomp} into the above relation yields
\begin{equation}
\label{eq:local_E_field}
\mathbf{E}(\mathbf{X})
=
\bar{\mathbf{E}}
-
\nabla \tilde{\phi}(\mathbf{X}),
\end{equation}
where $\bar{\mathbf{E}}$ denotes the macroscopic electric field and
$\tilde{\phi}(\mathbf{X})$ is the fluctuation potential.

Applying the volume average over the partitioned RVE domain gives
\begin{align}
\bar{\mathbf{E}}
&=
\frac{1}{V_\Omega} \int_\Omega \mathbf{E}(\mathbf{X}) \, d\Omega
\nonumber\\
&=
\sum_{\alpha=0}^{N_p-1} v_f^\alpha \bar{\mathbf{E}}
-
\sum_{\alpha=0}^{N_p-1}v_f^\alpha \left[ \frac{1}{V_\alpha}
\int_{\Omega_\alpha}
\nabla \tilde{\phi}(\mathbf{X}) \, d\Omega_\alpha \right]
\nonumber\\
&=
\sum_{\alpha=0}^{N_p-1} v_f^\alpha
\underbrace{\left[
\bar{\mathbf{E}}
-
\frac{1}{V_\alpha}
\int_{\Omega_\alpha}
\nabla \tilde{\phi}(\mathbf{X}) \, d\Omega_\alpha
\right]}_{\mathbf{E}^{\alpha}}.
\end{align}

Accordingly, the average electric field in the $\alpha$-th subdomain is
given by
\begin{equation}
\label{eq:subdomain_E_field}
\mathbf{E}^{\alpha}
=
\bar{\mathbf{E}}
-
\frac{1}{V_\alpha}
\int_{\Omega_\alpha}
\nabla \tilde{\phi}(\mathbf{X})\, d\Omega_\alpha .
\end{equation}

Applying the divergence theorem to the fluctuation terms in
Eqs.~\eqref{eq:subdomain_strain} and~\eqref{eq:subdomain_E_field}
transforms the subdomain volume integrals into boundary integrals. Under periodic boundary conditions (PBCs), the boundary of each subdomain $\Omega_\alpha$ comprises the outer boundary of the reduced microstructure and the internal interfaces $\Gamma^I_{\alpha\beta}$ shared with neighboring subdomains. Because the fluctuation fields are periodic, the outer-boundary contributions cancel, so that only the internal-interface contributions remain. The subdomain strain and electric field in $\Omega_\alpha$ are therefore given by
\begin{equation} \label{eq:subdomain_strain_divergence}
\boldsymbol{\varepsilon}^{\alpha}
=
\bar{\boldsymbol{\varepsilon}}
+
\frac{1}{V_\alpha}
\sum_{\beta=0}^{N_p-1}
\int_{\Gamma^I_{\alpha\beta}}
\boldsymbol{\xi}(\mathbf{X})
\otimes^{s}
\mathbf{N}^I_{\alpha\beta}
\, d\Gamma ,
\end{equation}

\begin{equation} \label{eq:subdomain_E_field_divergence}
\mathbf{E}^{\alpha}
=
\bar{\mathbf{E}}
-
\frac{1}{V_\alpha}
\sum_{\beta=0}^{N_p-1}
\int_{\Gamma^I_{\alpha\beta}}
\tilde{\phi}(\mathbf{X})
\, \mathbf{N}^I_{\alpha\beta}
\, d\Gamma .
\end{equation}
In the above expressions, $\mathbf{N}^I_{\alpha\beta}$ denotes the
outward unit normal vector on the interface
$\Gamma^I_{\alpha\beta}$ with respect to subdomain $\Omega_\alpha$.
If $\Omega_\alpha$ and $\Omega_\beta$ are not adjacent, then
$\Gamma^I_{\alpha\beta}=\varnothing$, and the corresponding term does
not contribute.

For notational convenience, we introduce the following interfacial interaction variables, where $\mathbf{a}_{\alpha\beta}$ is associated with displacement fluctuations and $b_{\alpha\beta}$ with electric-potential fluctuations:
\begin{equation}
\mathbf{a}_{\alpha\beta}
=
\frac{1}{V_\Omega}\int_{\Gamma^I_{\alpha\beta}}
\boldsymbol{\xi}(\mathbf{X})
\, d\Gamma ,
\end{equation}
and
\begin{equation}
b_{\alpha\beta}
=
\frac{1}{V_\Omega}\int_{\Gamma^I_{\alpha\beta}}
\tilde{\phi}(\mathbf{X})
\, d\Gamma .
\end{equation}

Using these definitions,
Eqs.~\eqref{eq:subdomain_strain_divergence}
and~\eqref{eq:subdomain_E_field_divergence}
can be written in the compact form

\begin{equation} \label{eq:subdomain_strain_compact}
\boldsymbol{\varepsilon}^{\alpha}
=
\bar{\boldsymbol{\varepsilon}}
+
\frac{V_\Omega}{V_\alpha}
\sum_{\beta=0}^{N_p-1}
\mathbf{a}_{\alpha\beta}
\otimes^{s}
\mathbf{N}^I_{\alpha\beta},
\end{equation}

\begin{equation} \label{eq:subdomain_E_field_compact}
\mathbf{E}^{\alpha}
=
\bar{\mathbf{E}}
-
\frac{V_\Omega}{V_\alpha}
\sum_{\beta=0}^{N_p-1}
b_{\alpha\beta}
\, \mathbf{N}^I_{\alpha\beta}.
\end{equation}

\subsection{Binary homogenization building block}

Building on the homogenization framework for reduced microstructures introduced in the previous subsection, we now derive the binary homogenization building block that serves as the basic unit of the PDMN for electromechanically coupled piezoelectricity. The key idea is to approximate the response of a complex RVE through a hierarchical assembly of two-subdomain interaction units, each governed by an analytical electromechanical homogenization relation.

\begin{figure}[htbp]
\centering
\includegraphics[width=1.0\linewidth]{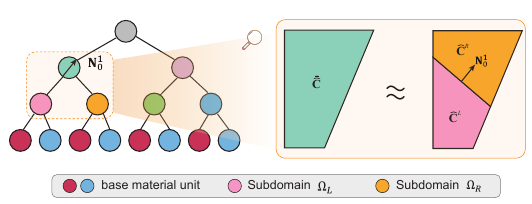}
\caption{Binary homogenization building block in the PDMN.
The unit cell consists of two subdomains $\Omega_L$ and $\Omega_R$
with generalized constitutive matrices $\widehat{\mathbf{C}}^L$ and $\widehat{\mathbf{C}}^R$, separated by
an interface with unit normal vector $\mathbf{N}$. Their interaction
produces the homogenized electromechanical constitutive matrix $\bar{\widehat{\mathbf{C}}}$.}
\label{fig:building_block}
\end{figure}
\newpage

As shown in Fig.~\ref{fig:building_block}, consider a unit cell $\Omega=\Omega_L\cup\Omega_R$ composed of two subdomains with volume fractions $v_f^L$ and $v_f^R$, separated by an interface of unit normal vector $\mathbf{N}$. For this binary partition, the compact subdomain relations introduced previously reduce to
\begin{equation}
\boldsymbol{\varepsilon}^{L}
=
\bar{\boldsymbol{\varepsilon}}
+
\frac{1}{v_f^{L}}\,\mathbf{a}\otimes^{s}\mathbf{N},
\qquad
\boldsymbol{\varepsilon}^{R}
=
\bar{\boldsymbol{\varepsilon}}
-
\frac{1}{v_f^{R}}\,\mathbf{a}\otimes^{s}\mathbf{N},
\end{equation}
and
\begin{equation}
\mathbf{E}^{L}
=
\bar{\mathbf{E}}
-
\frac{b}{v_f^{L}}\,\mathbf{N},
\qquad
\mathbf{E}^{R}
=
\bar{\mathbf{E}}
+
\frac{b}{v_f^{R}}\,\mathbf{N},
\end{equation}
where $\mathbf{a}$ and $b$ are the mechanical and electrical interaction variables associated with the interface.

To express these relations in a unified form, we introduce the generalized kinematic field and generalized flux vectors,
\begin{equation}
\widehat{\mathbf{z}}
=
\begin{bmatrix}
\boldsymbol{\varepsilon} \\
\mathbf{E}
\end{bmatrix},
\qquad
\widehat{\mathbf{y}}
=
\begin{bmatrix}
\boldsymbol{\sigma} \\
\mathbf{D}
\end{bmatrix},
\end{equation}
where $\boldsymbol{\varepsilon}$ and $\boldsymbol{\sigma}$ are written in Voigt notation. The local constitutive relation in each subdomain is then written in stress--charge form as
\begin{equation}
\widehat{\mathbf{y}}^{\alpha}
=
\widehat{\mathbf{C}}^{\alpha}\widehat{\mathbf{z}}^{\alpha},
\qquad
\widehat{\mathbf{C}}^{\alpha}
=
\begin{bmatrix}
\mathbf{C}^{E,\alpha} & -(\mathbf{e}^{\alpha})^{\mathsf{T}} \\
\mathbf{e}^{\alpha} & \boldsymbol{\kappa}^{\varepsilon,\alpha}
\end{bmatrix},
\qquad
\alpha\in\{L,R\},
\end{equation}
with $\mathbf{C}^{E,\alpha}$ denoting the elastic stiffness at constant electric field, $\mathbf{e}^{\alpha}$ the piezoelectric coupling matrix, and $\boldsymbol{\kappa}^{\varepsilon,\alpha}$ the dielectric permittivity at constant strain.

To explicitly account for the interface orientation, we define the operator
\begin{equation}
\mathbf{H}(\mathbf{N})=
\begin{bmatrix}
N_1 & 0   & 0 \\
0   & N_2 & 0 \\
0   & 0   & N_3 \\
0   & N_3 & N_2 \\
N_3 & 0   & N_1 \\
N_2 & N_1 & 0
\end{bmatrix},
\end{equation}
which provides the Voigt representation of the symmetric product $\mathbf{a}\otimes^s\mathbf{N}$.
Introducing the generalized interaction vector
$\widehat{\mathbf{x}}=[\,\mathbf{a}^{\mathsf T}\;\; b\,]^{\mathsf T}$ and the generalized localization operator
\begin{equation}
\widehat{\mathbf{A}}(\mathbf{N})
=
\begin{bmatrix}
\mathbf{H}(\mathbf{N}) & \mathbf{0}_{6\times 1} \\
\mathbf{0}_{3\times 3} & -\mathbf{N}
\end{bmatrix},
\end{equation}
The generalized fields in the two subdomains can be compactly expressed as
\begin{equation}
\widehat{\mathbf{z}}^{L}
=
\bar{\widehat{\mathbf{z}}}
+
\frac{1}{v_f^{L}}\,\widehat{\mathbf{A}}(\mathbf{N})\widehat{\mathbf{x}},
\qquad
\widehat{\mathbf{z}}^{R}
=
\bar{\widehat{\mathbf{z}}}
-
\frac{1}{v_f^{R}}\,\widehat{\mathbf{A}}(\mathbf{N})\widehat{\mathbf{x}},
\end{equation}
where $\bar{\widehat{\mathbf{z}}}=[\,\bar{\boldsymbol{\varepsilon}}^{\mathsf T}\;\;\bar{\mathbf{E}}^{\mathsf T}\,]^{\mathsf T}$.

The interaction variables are determined from the interfacial equilibrium. Specifically, continuity of traction and of the normal electric displacement is enforced through
\begin{equation}
\widehat{\mathbf{B}}(\mathbf{N})\,\widehat{\mathbf{y}}^{L}
=
\widehat{\mathbf{B}}(\mathbf{N})\,\widehat{\mathbf{y}}^{R},
\end{equation}
with the interface operator
\begin{equation}
\widehat{\mathbf{B}}(\mathbf{N})
=
\begin{bmatrix}
\mathbf{H}(\mathbf{N})^{\mathsf{T}} & \mathbf{0}_{3\times 3} \\
\mathbf{0}_{1\times 6} & \mathbf{N}^{\mathsf{T}}
\end{bmatrix}.
\end{equation}
Substituting the constitutive and localization relations into the above condition yields a linear system for $\widehat{\mathbf{x}}$,
\begin{equation}
\widehat{\mathbf{S}}\,\widehat{\mathbf{x}}
=
-
\widehat{\mathbf{B}}(\mathbf{N})
\left(
\widehat{\mathbf{C}}^{L}
-
\widehat{\mathbf{C}}^{R}
\right)\bar{\widehat{\mathbf{z}}},
\end{equation}
where
\begin{equation}
\widehat{\mathbf{S}}
=
\widehat{\mathbf{B}}(\mathbf{N})
\left(
\frac{1}{v_f^{L}}\widehat{\mathbf{C}}^{L}
+
\frac{1}{v_f^{R}}\widehat{\mathbf{C}}^{R}
\right)
\widehat{\mathbf{A}}(\mathbf{N}).
\end{equation}

After eliminating $\widehat{\mathbf{x}}$ and substituting the result into the subdomain averaging relation
$\bar{\widehat{\mathbf{y}}}=v_f^L\widehat{\mathbf{y}}^L+v_f^R\widehat{\mathbf{y}}^R$, the homogenized constitutive response of the binary unit can be written as
$\bar{\widehat{\mathbf{y}}}=\bar{\widehat{\mathbf{C}}}\,\bar{\widehat{\mathbf{z}}}$,
where the effective generalized constitutive matrix is given in closed form by
\begin{equation}\label{eq:H2_function}
\bar{\widehat{\mathbf{C}}}
=
\widehat{\mathbb{H}}_{2}
\left(
\widehat{\mathbf{C}}^{L},
\widehat{\mathbf{C}}^{R},
\mathbf{N}, v_f^{L}, v_f^{R}
\right)
=
v_f^{L}\widehat{\mathbf{C}}^{L}
+
v_f^{R}\widehat{\mathbf{C}}^{R}
-
v_f^{L}v_f^{R}
\left(
\widehat{\mathbf{C}}^{L}
-
\widehat{\mathbf{C}}^{R}
\right)
\widehat{\mathbf{Q}}
\left(
\widehat{\mathbf{C}}^{L}
-
\widehat{\mathbf{C}}^{R}
\right),
\end{equation}
where
\begin{equation}\label{eq:Q_piezo}
\widehat{\mathbf{Q}}
=
\widehat{\mathbf{A}}(\mathbf{N})\,
\widetilde{\mathbf{S}}^{-1}\,
\widehat{\mathbf{B}}(\mathbf{N}),
\qquad
\widetilde{\mathbf{S}}
=
\widehat{\mathbf{B}}(\mathbf{N})
\left(
v_f^{R}\widehat{\mathbf{C}}^{L}
+
v_f^{L}\widehat{\mathbf{C}}^{R}
\right)
\widehat{\mathbf{A}}(\mathbf{N}).
\end{equation}

This analytical operator constitutes the fundamental binary homogenization unit of the PDMN. Applying it recursively across the network hierarchy yields the effective electromechanical constitutive response of the reduced microstructure.

\subsection{PDMN formulation}
The effective response of the original RVE is approximated by a reduced microstructure that provides a simplified representation of the underlying heterogeneous morphology. In the proposed framework, the topology of this reduced microstructure is parameterized by the PDMN through a recursive binary partition: after $N$ successive splits, the reduced microstructure is decomposed into $2^{N}$ non-overlapping terminal subdomains. This hierarchical partition defines the architecture of the PDMN, as illustrated in Fig.~\ref{fig:Fig2_reducedMicrostructure}.

The PDMN consists of two components. The first is a binary \emph{material network} that encodes the hierarchical topology of the reduced microstructure. Each internal node $(d,p)$ in the tree is characterized by an interface orientation, represented by a unit normal vector $\mathbf{N}^{d}_{p}$, where $d$ denotes the depth of the hierarchy and $p$ the node index at that depth. This unit normal is parameterized by two trainable angles, $\theta^{d}_{p}$ and $\phi^{d}_{p}$, such that
\begin{equation}\label{eq:direction_vector}
\mathbf{N}^{d}_{p} =
\begin{bmatrix}
\cos(2\pi\phi^{d}_{p}) \sin(\pi \theta^{d}_{p}) \\
\sin(2\pi\phi^{d}_{p}) \sin(\pi \theta^{d}_{p}) \\
\cos(\pi \theta^{d}_{p})
\end{bmatrix}.
\end{equation}

The second component is a set of \emph{material nodes}, each corresponding to a terminal subdomain and its constituent material. Each material node is parameterized by a trainable scalar $z^{i}$, from which a positive weight is obtained through the activation function
\begin{equation}\label{eq:activation_function}
W^{i} = \operatorname{softplus}(z^{i})
= \ln\left(1+\exp(z^{i})\right).
\end{equation}

At the deepest level of the hierarchy, each material node $\mathcal{M}^{i}$ corresponds to a terminal subdomain $\Omega^{N}_{i}$, so that the reduced microstructure domain can be written as
\begin{equation}
\Omega = \bigcup_{i=0}^{2^{N}-1} \Omega^{N}_{i}.
\end{equation}
The corresponding volume fraction is defined as
\begin{equation}
v_f^{i} = \frac{W^{i}}{\sum_{j=0}^{2^{N}-1} W^{j}}.
\end{equation}

For the two-phase setting considered in this study, the material nodes are assigned according to index parity: nodes with even indices, $i=0,2,\ldots,2^{N}-2$, belong to phase~1, whereas nodes with odd indices, $i=1,3,\ldots,2^{N}-1$, belong to phase~2.

At each node $(d,p)$ of the material network, the corresponding binary partition divides the parent subdomain $\Omega^{d}_{p}$ into two child subdomains, $\Omega^{d+1}_{2p}$ and $\Omega^{d+1}_{2p+1}$. To represent the terminal material nodes contained in these two branches, we introduce two disjoint sets, $\mathcal{J}^{0}_{d,p}$ and $\mathcal{J}^{1}_{d,p}$, which collect the terminal descendants in the left and right branches, respectively. These sets are recursively defined as
\begin{equation}\label{eq:J_defined}
\mathcal{J}^{0}_{d,p} =
\begin{cases}
\mathcal{J}^{0}_{d+1,2p} \cup \mathcal{J}^{1}_{d+1,2p}, & d < N-1, \\
\left\{ \mathcal{M}^{2p} \right\}, & d = N-1,
\end{cases}
\end{equation}
and
\begin{equation}
\mathcal{J}^{1}_{d,p} =
\begin{cases}
\mathcal{J}^{0}_{d+1,2p+1} \cup \mathcal{J}^{1}_{d+1,2p+1}, & d < N-1, \\
\left\{ \mathcal{M}^{2p+1} \right\}, & d = N-1.
\end{cases}
\end{equation}

Accordingly, $\mathcal{J}^{0}_{d,p}$ and $\mathcal{J}^{1}_{d,p}$ identify the terminal material nodes contained in the two child subdomains generated by the partition at node $(d,p)$. These sets are introduced to define the recursive localization relations throughout the hierarchy.

\subsubsection{Field localization in the PDMN}

Within the PDMN framework, the strain and electric fields in each child subdomain are obtained recursively from those in the corresponding parent subdomain. At a given depth $d = 0,\ldots,N-1$ and node index $p = 0,\ldots,2^{d}-1$, the parent subdomain $\Omega^{d}_{p}$ is partitioned into two child subdomains, $\Omega^{d+1}_{2p}$ and $\Omega^{d+1}_{2p+1}$, whose sets of terminal material nodes are denoted by $\mathcal{J}^{0}_{d,p}$ and $\mathcal{J}^{1}_{d,p}$, respectively.

The strain fields in the two child subdomains are given by
\begin{equation}
\label{eq:strain_localization_block}
\boldsymbol{\varepsilon}^{d+1}_{q}
=
\boldsymbol{\varepsilon}^{d}_{p}
+
\begin{cases}
\displaystyle
\frac{V_{\Omega}}
     {V_{\mathcal{J}^{0}_{d,p}}}
\,\mathbf{a}^{d}_{p} \otimes^{s} \mathbf{N}^{d}_{p},
& q = 2p, \\
\displaystyle
-\frac{V_{\Omega}}
     {V_{\mathcal{J}^{1}_{d,p}}}
\,\mathbf{a}^{d}_{p} \otimes^{s} \mathbf{N}^{d}_{p},
& q = 2p+1 .
\end{cases}
\end{equation}

Analogously, the electric fields in the two child subdomains are given by
\begin{equation}
\label{eq:electric_localization_block}
\mathbf{E}^{d+1}_{q}
=
\mathbf{E}^{d}_{p}
-
\begin{cases}
\displaystyle
\frac{V_{\Omega}}
     {V_{\mathcal{J}^{0}_{d,p}}}
\,b^{d}_{p}\,\mathbf{N}^{d}_{p},
& q = 2p, \\
\displaystyle
-\frac{V_{\Omega}}
     {V_{\mathcal{J}^{1}_{d,p}}}
\,b^{d}_{p}\,\mathbf{N}^{d}_{p},
& q = 2p+1 .
\end{cases}
\end{equation}
Here, $V_{\mathcal{J}^{0}_{d,p}}$ and $V_{\mathcal{J}^{1}_{d,p}}$ denote the total volumes of the terminal descendants contained in the two branches of node $(d,p)$.

At the root node, the strain and electric fields are prescribed by the macroscopic fields,
\begin{equation}
\boldsymbol{\varepsilon}^{0}_{0} = \bar{\boldsymbol{\varepsilon}},
\qquad
\mathbf{E}^{0}_{0} = \bar{\mathbf{E}}.
\end{equation}

By recursively propagating the block-wise localization relations through the hierarchy, the strain and electric fields in a terminal subdomain $\Omega^{N}_{i}$ can be expressed explicitly in terms of the macroscopic driving fields and the interaction variables distributed throughout the network. The resulting localization mappings read
\begin{align}
\label{eq:interaction_mapping_strain}
\boldsymbol{\varepsilon}^{i}
&=
\bar{\boldsymbol{\varepsilon}}
+
\sum_{d=0}^{N-1}
\sum_{p=0}^{2^{d}-1}
\alpha^{i}_{d,p}\,
\mathbf{a}^{d}_{p}
\otimes^{s}
\mathbf{N}^{d}_{p}, \\
\label{eq:interaction_mapping_electric}
\mathbf{E}^{i}
&=
\bar{\mathbf{E}}
-
\sum_{d=0}^{N-1}
\sum_{p=0}^{2^{d}-1}
\alpha^{i}_{d,p}\,
b^{d}_{p}\,
\mathbf{N}^{d}_{p},
\end{align}
where the coefficients $\alpha^{i}_{d,p}$ encode the hierarchical topology of the PDMN and quantify the contribution of the interaction at node $(d,p)$ to the terminal subdomain $\Omega_i^N$.

Specifically, the localization coefficients are defined as
\begin{equation}
\label{eq:alpha_defined}
\alpha^{i}_{d,p}
=
\begin{cases}
\displaystyle
\frac{1}{\sum\limits_{k\in\mathcal{J}^{0}_{d,p}} v_f^{k}},
& i\in\mathcal{J}^{0}_{d,p}, \\
\displaystyle
-\frac{1}{\sum\limits_{k\in\mathcal{J}^{1}_{d,p}} v_f^{k}},
& i\in\mathcal{J}^{1}_{d,p}, \\
0, & \text{otherwise},
\end{cases}
\end{equation}
where $\mathcal{J}^{0}_{d,p}$ and $\mathcal{J}^{1}_{d,p}$ denote the sets of material nodes located on the two sides of the interface associated with node $(d,p)$.

\subsubsection{Hill--Mandel consistency and interaction equilibrium}

Using the localization relations Eqs.~\eqref{eq:interaction_mapping_strain} and~\eqref{eq:interaction_mapping_electric}, the Hill--Mandel conditions Eqs.~\eqref{eq:mech_hill_mandel} and~\eqref{eq:elec_hill_mandel} can be rewritten in terms of the PDMN interaction variables.

Substituting Eq.~\eqref{eq:interaction_mapping_strain} into the mechanical Hill--Mandel condition gives
\begin{equation}
\label{eq:mech_hill_mandel_PDMN}
\bar{\boldsymbol{\sigma}} : \delta \bar{\boldsymbol{\varepsilon}}
=
\sum_{i=0}^{2^{N}-1}
v_f^{i}\,
\boldsymbol{\sigma}^{i}
:
\left(
\delta \bar{\boldsymbol{\varepsilon}}
+
\sum_{d=0}^{N-1}
\sum_{p=0}^{2^{d}-1}
\alpha^{i}_{d,p}\,
\delta \mathbf{a}^{d}_{p}
\otimes^{s}
\mathbf{N}^{d}_{p}
\right).
\end{equation}
Similarly, substitution of Eq.~\eqref{eq:interaction_mapping_electric} into the electric Hill--Mandel condition yields
\begin{equation}
\label{eq:elec_hill_mandel_PDMN}
\bar{\mathbf{D}} \cdot \delta \bar{\mathbf{E}}
=
\sum_{i=0}^{2^{N}-1}
v_f^{i}\,
\mathbf{D}^{i}
\cdot
\left(
\delta \bar{\mathbf{E}}
-
\sum_{d=0}^{N-1}
\sum_{p=0}^{2^{d}-1}
\alpha^{i}_{d,p}\,
\delta b^{d}_{p}\,
\mathbf{N}^{d}_{p}
\right).
\end{equation}

Because the macroscopic variations
$\delta \bar{\boldsymbol{\varepsilon}}$ and $\delta \bar{\mathbf{E}}$
are independent of the interaction variations
$\delta \mathbf{a}^{d}_{p}$ and $\delta b^{d}_{p}$, the Hill--Mandel principle requires the microscopic power contribution associated with each admissible interaction variation to vanish independently. 

For the mechanical field, this condition takes the form
\begin{equation}
\label{eq:interaction_equilibrium_mech}
\sum_{i=0}^{2^{N}-1}
v_f^{i}\,
\alpha^{i}_{d,p}\,
\boldsymbol{\sigma}^{i}
\cdot
\mathbf{N}^{d}_{p}
=
\mathbf{0},
\qquad
d = 0,\ldots,N-1,\;
p = 0,\ldots,2^{d}-1.
\end{equation}
For the electric field, one analogously obtains
\begin{equation}
\label{eq:interaction_equilibrium_elec}
\sum_{i=0}^{2^{N}-1}
v_f^{i}\,
\alpha^{i}_{d,p}\,
\mathbf{D}^{i}
\cdot
\mathbf{N}^{d}_{p}
=
0,
\qquad
d = 0,\ldots,N-1,\;
p = 0,\ldots,2^{d}-1.
\end{equation}

The local constitutive response of each terminal subdomain is expressed in the general form
\begin{equation}
\label{eq:localLaw1}
\boldsymbol{\sigma}^{i}
=
\mathcal{D}
\bigl(
\boldsymbol{\varepsilon}^{i},\,
z^{i}_{\varepsilon},\,
\mathbf{E}^{i},\,
z^{i}_{E}
\bigr),
\end{equation}
and
\begin{equation}
\label{eq:localLaw2}
\mathbf{D}^{i}
=
\mathcal{P}
\bigl(
\boldsymbol{\varepsilon}^{i},\,
z^{i}_{\varepsilon},\,
\mathbf{E}^{i},\,
z^{i}_{E}
\bigr),
\end{equation}
where $z^{i}_{\varepsilon}$ and $z^{i}_{E}$ denote the internal variables associated with the mechanical and electric responses, respectively. The associated consistent local tangents are denoted $\mathbf{C}_{\sigma\varepsilon}^{i}=\partial\boldsymbol{\sigma}^{i}/\partial\boldsymbol{\varepsilon}^{i}$, $\mathbf{C}_{\sigma E}^{i}=\partial\boldsymbol{\sigma}^{i}/\partial\mathbf{E}^{i}$, $\mathbf{C}_{D\varepsilon}^{i}=\partial\mathbf{D}^{i}/\partial\boldsymbol{\varepsilon}^{i}$, and $\mathbf{C}_{DE}^{i}=\partial\mathbf{D}^{i}/\partial\mathbf{E}^{i}$.

The interaction variables $\mathbf{a}^{d}_{p}$ and $b^{d}_{p}$ are determined by enforcing Eqs.~\eqref{eq:interaction_equilibrium_mech} and~\eqref{eq:interaction_equilibrium_elec}. Accordingly, the residuals at each internal node are defined as
\begin{equation}
\label{eq:residual_mech}
\mathbf{r}_{\mathrm{mech}}^{\,d,p}
=
\sum_{i=0}^{2^{N}-1}
v_f^{i}\,
\alpha^{i}_{d,p}\,
\boldsymbol{\sigma}^{i}
\cdot
\mathbf{N}^{d}_{p},
\qquad
d = 0,\ldots,N-1,\;
p = 0,\ldots,2^{d}-1,
\end{equation}
and
\begin{equation}
\label{eq:residual_elec}
r_{\mathrm{elec}}^{\,d,p}
=
\sum_{i=0}^{2^{N}-1}
v_f^{i}\,
\alpha^{i}_{d,p}\,
\mathbf{D}^{i}
\cdot
\mathbf{N}^{d}_{p},
\qquad
d = 0,\ldots,N-1,\;
p = 0,\ldots,2^{d}-1.
\end{equation}

Collecting the residual equations over all internal nodes, the PDMN yields $3(2^{N}-1)$ mechanical equations and $(2^{N}-1)$ electric equations. These coupled nonlinear equations determine the interaction variables through the conditions $\mathbf{r}_{\mathrm{mech}}^{\,d,p}=\mathbf{0}$ and $r_{\mathrm{elec}}^{\,d,p}=0$ for all internal nodes. Their numerical solution is developed in the following subsection by a fully coupled Newton--Raphson scheme.

\subsubsection{Coupled Newton--Raphson scheme for interaction variables}

To solve the coupled residual equations
Eqs.~\eqref{eq:residual_mech} and~\eqref{eq:residual_elec}, all interaction variables are assembled into a global vector.
This representation yields a compact formulation and facilitates the derivation of a fully coupled Newton--Raphson scheme.

To this end, each internal node $(d,p)$ of the material network is assigned a unique global index
\begin{equation}
k(d,p)
=
\sum_{m=0}^{d-1}2^{m}+p
=
(2^{d}-1)+p,
\qquad
k=0,\ldots,2^{N}-2.
\end{equation}

Using this indexing, the vector-valued mechanical interaction variables
$\mathbf{a}^{d}_{p}\in\mathbb{R}^{3}$ are stacked into the global vector
$\mathbf{A}\in\mathbb{R}^{3(2^{N}-1)}$ according to
\begin{equation}
A_{3k(d,p)+j}
=
a^{d}_{p,j},
\qquad
j=1,2,3,
\end{equation}
while the scalar electric interaction variables $b^{d}_{p}$ are assembled into
$\mathbf{B}\in\mathbb{R}^{2^{N}-1}$ through
\begin{equation}
B_{k(d,p)}
=
b^{d}_{p}.
\end{equation}

With the interaction variables written in global form, the localization
relations in
Eqs.~\eqref{eq:interaction_mapping_strain}
and~\eqref{eq:interaction_mapping_electric}
can be expressed compactly as
\begin{equation}
\label{eq:interaction_mapping_strain_vec}
\mathrm{vec}^{\boldsymbol{\varepsilon}}
\!\left(\boldsymbol{\varepsilon}^{i}\right)
=
\mathrm{vec}^{\boldsymbol{\varepsilon}}
\!\left(\bar{\boldsymbol{\varepsilon}}\right)
+
\mathbf{M}^{i}\mathbf{A},
\end{equation}
and

\begin{equation}
\label{eq:interaction_mapping_electric_vec}
\mathbf{E}^{i}
=
\bar{\mathbf{E}}
-
\mathbf{O}^{i}\mathbf{B}.
\end{equation}
Here,
$\mathrm{vec}^{\boldsymbol{\varepsilon}}(\cdot)$ and
$\mathrm{vec}^{\boldsymbol{\sigma}}(\cdot)$ denote the Voigt
representations of strain and stress, respectively:
\[
\mathrm{vec}^{\boldsymbol{\varepsilon}}
\!\left(
\begin{bmatrix}
\varepsilon_{11} & \varepsilon_{12} & \varepsilon_{13}\\
\varepsilon_{12} & \varepsilon_{22} & \varepsilon_{23}\\
\varepsilon_{13} & \varepsilon_{23} & \varepsilon_{33}
\end{bmatrix}
\right)
=
\begin{bmatrix}
\varepsilon_{11}\\
\varepsilon_{22}\\
\varepsilon_{33}\\
2\varepsilon_{23}\\
2\varepsilon_{13}\\
2\varepsilon_{12}
\end{bmatrix},
\qquad
\mathrm{vec}^{\boldsymbol{\sigma}}
\!\left(
\begin{bmatrix}
\sigma_{11} & \sigma_{12} & \sigma_{13}\\
\sigma_{12} & \sigma_{22} & \sigma_{23}\\
\sigma_{13} & \sigma_{23} & \sigma_{33}
\end{bmatrix}
\right)
=
\begin{bmatrix}
\sigma_{11}\\
\sigma_{22}\\
\sigma_{33}\\
\sigma_{23}\\
\sigma_{13}\\
\sigma_{12}
\end{bmatrix}.
\]

Conversely, $\mathrm{mat}(\cdot)$ denotes the inverse of the $\mathrm{vec}(\cdot)$ operation, returning the $6\times 6$ Voigt matrix representation of a fourth-order tensor; it is used for the local and homogenized tangents introduced below.

The matrices $\mathbf{M}^{i}\in\mathbb{R}^{6\times 3(2^{N}-1)}$ and $\mathbf{O}^{i}\in\mathbb{R}^{3\times(2^{N}-1)}$ assemble the contributions of all interaction variables associated with the terminal subdomain $i$. They are defined as
\begin{equation}
\mathbf{M}^{i}
=
\operatorname{row}_{(d,p)}
\Big(
\alpha^{i}_{d,p}\,\mathbf{H}(\mathbf{N}^{d}_{p})
\Big),
\end{equation}
where $\operatorname{row}_{(d,p)}(\cdot)$ denotes block-row concatenation over all interaction pairs $(d,p)$. Likewise,

\begin{equation}
\mathbf{O}^{i}
=
\operatorname{row}_{(d,p)}
\Big(
\alpha^{i}_{d,p}\,\mathbf{N}^{d}_{p}
\Big).
\end{equation}

Substituting the matrices
$\mathbf{M}^{i}$ and
$\mathbf{O}^{i}$
into the node-wise residual equations yields the global residual vectors
\begin{equation}
\label{eq:r_mech}
\mathbf{r}_{\mathrm{mech}}
=
\sum_{i=0}^{2^{N}-1}
v_f^{i}\,
(\mathbf{M}^{i})^{\mathsf{T}}\,
\mathrm{vec}^{\boldsymbol{\sigma}}
\!\left(\boldsymbol{\sigma}^{i}\right),
\qquad
\mathbf{r}_{\mathrm{mech}}\in\mathbb{R}^{3(2^{N}-1)},
\end{equation}
and
\begin{equation}
\label{eq:r_elec}
\mathbf{r}_{\mathrm{elec}}
=
\sum_{i=0}^{2^{N}-1}
v_f^{i}\,
(\mathbf{O}^{i})^{\mathsf{T}}\,
\mathbf{D}^{i},
\qquad
\mathbf{r}_{\mathrm{elec}}\in\mathbb{R}^{2^{N}-1}.
\end{equation}

The coupled electromechanical problem is solved by a fully coupled
Newton--Raphson scheme. At iteration $k$, linearization of the global residual equations gives
\begin{equation}
\label{eq:coupled_newton}
\begin{bmatrix}
\dfrac{\partial \mathbf{r}_{\mathrm{mech}}}{\partial \mathbf{A}} &
\dfrac{\partial \mathbf{r}_{\mathrm{mech}}}{\partial \mathbf{B}} \\
\dfrac{\partial \mathbf{r}_{\mathrm{elec}}}{\partial \mathbf{A}} &
\dfrac{\partial \mathbf{r}_{\mathrm{elec}}}{\partial \mathbf{B}}
\end{bmatrix}^{(k)}
\begin{bmatrix}
\Delta \mathbf{A}^{(k)} \\
\Delta \mathbf{B}^{(k)}
\end{bmatrix}
=
-
\begin{bmatrix}
\mathbf{r}_{\mathrm{mech}}^{(k)} \\
\mathbf{r}_{\mathrm{elec}}^{(k)}
\end{bmatrix}.
\end{equation}
The interaction variables are then updated according to
\begin{equation}
\mathbf{A}^{(k+1)}
=
\mathbf{A}^{(k)}+\Delta \mathbf{A}^{(k)},
\qquad
\mathbf{B}^{(k+1)}
=
\mathbf{B}^{(k)}+\Delta \mathbf{B}^{(k)}.
\end{equation}

The Jacobian blocks appearing in Eq.~\eqref{eq:coupled_newton} are given by
\begin{equation}
\label{eq:drmech_dA}
\frac{\partial \mathbf{r}_{\mathrm{mech}}}{\partial \mathbf{A}}
=
\sum_{i=0}^{2^{N}-1}
v_f^{i}\,
(\mathbf{M}^{i})^{\mathsf{T}}\,
\mathrm{mat}
\!\left(
\frac{\partial \boldsymbol{\sigma}^{i}}
     {\partial \boldsymbol{\varepsilon}^{i}}
\right)
\mathbf{M}^{i},
\end{equation}

\begin{equation}
\label{eq:drmech_dB}
\frac{\partial \mathbf{r}_{\mathrm{mech}}}{\partial \mathbf{B}}
=
-\sum_{i=0}^{2^{N}-1}
v_f^{i}\,
(\mathbf{M}^{i})^{\mathsf{T}}\,
\mathrm{mat}
\!\left(
\frac{\partial \boldsymbol{\sigma}^{i}}
     {\partial \mathbf{E}^{i}}
\right)
\mathbf{O}^{i},
\end{equation}

\begin{equation}
\label{eq:drelec_dA}
\frac{\partial \mathbf{r}_{\mathrm{elec}}}{\partial \mathbf{A}}
=
\sum_{i=0}^{2^{N}-1}
v_f^{i}\,
(\mathbf{O}^{i})^{\mathsf{T}}\,
\frac{\partial \mathbf{D}^{i}}
     {\partial \boldsymbol{\varepsilon}^{i}}
\mathbf{M}^{i},
\end{equation}

\begin{equation}
\label{eq:drelec_dB}
\frac{\partial \mathbf{r}_{\mathrm{elec}}}{\partial \mathbf{B}}
=
-\sum_{i=0}^{2^{N}-1}
v_f^{i}\,
(\mathbf{O}^{i})^{\mathsf{T}}\,
\frac{\partial \mathbf{D}^{i}}
     {\partial \mathbf{E}^{i}}
\mathbf{O}^{i}.
\end{equation}

\subsubsection{Homogenized response and consistent tangent operators}

Once the Hill--Mandel conditions have converged, the homogenized stress and electric displacement are recovered by volume averaging over all terminal subdomains, namely,
\begin{equation}
\label{eq:macro_sigma_avg}
\mathrm{vec}^{\boldsymbol{\sigma}}(\bar{\boldsymbol{\sigma}})
=
\sum_{i=0}^{2^{N}-1}
v_f^{i}\,
\mathrm{vec}^{\boldsymbol{\sigma}}(\boldsymbol{\sigma}^{i}),
\qquad
\bar{\mathbf{D}}
=
\sum_{i=0}^{2^{N}-1}
v_f^{i}\,\mathbf{D}^{i}.
\end{equation}

Introducing the generalized homogenized response and driving field vectors as
\begin{equation}
\label{eq:macro_generalized_yz}
\bar{\widehat{\mathbf y}}
=
\begin{bmatrix}
\mathrm{vec}^{\boldsymbol{\sigma}}(\bar{\boldsymbol{\sigma}})\\
\bar{\mathbf D}
\end{bmatrix},
\qquad
\bar{\widehat{\mathbf z}}
=
\begin{bmatrix}
\mathrm{vec}^{\boldsymbol{\varepsilon}}(\bar{\boldsymbol{\varepsilon}})\\
\bar{\mathbf E}
\end{bmatrix},
\end{equation}
the consistent homogenized tangent operator $\widehat{\mathbf C}^{\mathrm{alg}}_{\mathrm{hom}}$, where the superscript ``alg'' stands for algorithmic, is defined by
\begin{equation}
\label{eq:Chat_general}
\widehat{\mathbf C}^{\mathrm{alg}}_{\mathrm{hom}}
=
\frac{\partial \bar{\widehat{\mathbf y}}}
{\partial \bar{\widehat{\mathbf z}}}
=
\begin{bmatrix}
\mathrm{mat}
\!\left(
\frac{\partial \bar{\boldsymbol{\sigma}}}
     {\partial \bar{\boldsymbol{\varepsilon}}}
\right)
&
\dfrac{\partial \,\mathrm{vec}^{\boldsymbol{\sigma}}(\bar{\boldsymbol{\sigma}})}
      {\partial \bar{\mathbf E}}
\\
\dfrac{\partial \bar{\mathbf D}}
      {\partial \,\mathrm{vec}^{\boldsymbol{\varepsilon}}(\bar{\boldsymbol{\varepsilon}})}
&
\dfrac{\partial \bar{\mathbf D}}
      {\partial \bar{\mathbf E}}
\end{bmatrix}.
\end{equation}

Combining the relations with the locally consistent electromechanical tangent operators, the four block components of
$\widehat{\mathbf C}^{\mathrm{alg}}_{\mathrm{hom}}$ are given by

\begin{align}
\label{eq:Chat_sig_eps}
\mathrm{mat}
\!\left(
\frac{\partial \bar{\boldsymbol{\sigma}}}
     {\partial \bar{\boldsymbol{\varepsilon}}}
\right)
&=
\sum_{i=0}^{2^N-1}
v_f^i\,
\mathrm{mat}
\!\left(
\frac{\partial \boldsymbol{\sigma}^{i}}
     {\partial \boldsymbol{\varepsilon}^{i}}
\right)
\nonumber\\
&\quad
+
\sum_{i=0}^{2^N-1}
v_f^i\,
\mathrm{mat}
\!\left(
\frac{\partial \boldsymbol{\sigma}^{i}}
     {\partial \boldsymbol{\varepsilon}^{i}}
\right)
\mathbf M^i
\frac{\partial \mathbf A}
     {\partial \,\mathrm{vec}^{\boldsymbol{\varepsilon}}(\bar{\boldsymbol{\varepsilon}})}
\nonumber\\
&\quad
-
\sum_{i=0}^{2^N-1}
v_f^i\,
\frac{\partial \boldsymbol{\sigma}^{i}}
     {\partial \mathbf E^{i}}
\mathbf O^i
\frac{\partial \mathbf B}
     {\partial \,\mathrm{vec}^{\boldsymbol{\varepsilon}}(\bar{\boldsymbol{\varepsilon}})},
\\
\label{eq:Chat_sig_E}
\frac{\partial \,\mathrm{vec}^{\boldsymbol{\sigma}}(\bar{\boldsymbol{\sigma}})}
     {\partial \bar{\mathbf E}}
&=
\sum_{i=0}^{2^N-1}
v_f^i\,
\mathrm{mat}
\!\left(
\frac{\partial \boldsymbol{\sigma}^{i}}
     {\partial \boldsymbol{\varepsilon}^{i}}
\right)
\mathbf M^i
\frac{\partial \mathbf A}{\partial \bar{\mathbf E}}
\nonumber\\
&\quad
+
\sum_{i=0}^{2^N-1}
v_f^i\,
\frac{\partial \boldsymbol{\sigma}^{i}}
     {\partial \mathbf E^{i}}
\nonumber\\
&\quad
-
\sum_{i=0}^{2^N-1}
v_f^i\,
\frac{\partial \boldsymbol{\sigma}^{i}}
     {\partial \mathbf E^{i}}
\mathbf O^i
\frac{\partial \mathbf B}{\partial \bar{\mathbf E}},
\\
\label{eq:Chat_D_eps}
\frac{\partial \bar{\mathbf D}}
     {\partial \,\mathrm{vec}^{\boldsymbol{\varepsilon}}(\bar{\boldsymbol{\varepsilon}})}
&=
\sum_{i=0}^{2^N-1}
v_f^i\,
\frac{\partial \mathbf D^i}
     {\partial \boldsymbol{\varepsilon}^{i}}
\nonumber\\
&\quad
+
\sum_{i=0}^{2^N-1}
v_f^i\,
\frac{\partial \mathbf D^i}
     {\partial \boldsymbol{\varepsilon}^{i}}
\mathbf M^i
\frac{\partial \mathbf A}
     {\partial \,\mathrm{vec}^{\boldsymbol{\varepsilon}}(\bar{\boldsymbol{\varepsilon}})}
\nonumber\\
&\quad
-
\sum_{i=0}^{2^N-1}
v_f^i\,
\frac{\partial \mathbf D^i}
     {\partial \mathbf E^{i}}
\mathbf O^i
\frac{\partial \mathbf B}
     {\partial \,\mathrm{vec}^{\boldsymbol{\varepsilon}}(\bar{\boldsymbol{\varepsilon}})},
\\
\label{eq:Chat_D_E}
\frac{\partial \bar{\mathbf D}}
     {\partial \bar{\mathbf E}}
&=
\sum_{i=0}^{2^N-1}
v_f^i\,
\frac{\partial \mathbf D^i}
     {\partial \boldsymbol{\varepsilon}^{i}}
\mathbf M^i
\frac{\partial \mathbf A}{\partial \bar{\mathbf E}}
\nonumber\\
&\quad
+
\sum_{i=0}^{2^N-1}
v_f^i\,
\frac{\partial \mathbf D^i}
     {\partial \mathbf E^{i}}
\nonumber\\
&\quad
-
\sum_{i=0}^{2^N-1}
v_f^i\,
\frac{\partial \mathbf D^i}
     {\partial \mathbf E^{i}}
\mathbf O^i
\frac{\partial \mathbf B}{\partial \bar{\mathbf E}}.
\end{align}

The required sensitivities of the interaction variables are obtained by differentiating the coupled residual equations with respect to the macroscopic driving fields. With the coupled Jacobian defined as

\begin{equation}
\label{eq:coupled_jacobian_sensitivity}
\mathbf J
=
\begin{bmatrix}
\dfrac{\partial \mathbf r_{\mathrm{mech}}}{\partial \mathbf A} &
\dfrac{\partial \mathbf r_{\mathrm{mech}}}{\partial \mathbf B}
\\[6pt]
\dfrac{\partial \mathbf r_{\mathrm{elec}}}{\partial \mathbf A} &
\dfrac{\partial \mathbf r_{\mathrm{elec}}}{\partial \mathbf B}
\end{bmatrix},
\end{equation}
The sensitivities with respect to the macroscopic strain and electric field are obtained as
\begin{align}
\label{eq:sensitivity_eps_compact_inverse}
\begin{bmatrix}
\dfrac{\partial \mathbf A}
      {\partial \,\mathrm{vec}^{\boldsymbol{\varepsilon}}(\bar{\boldsymbol{\varepsilon}})}
\\[8pt]
\dfrac{\partial \mathbf B}
      {\partial \,\mathrm{vec}^{\boldsymbol{\varepsilon}}(\bar{\boldsymbol{\varepsilon}})}
\end{bmatrix}
&=
-
\mathbf J^{-1}
\begin{bmatrix}
\displaystyle
\sum_{i=0}^{2^N-1}
v_f^i (\mathbf M^i)^{\mathsf T}
\mathrm{mat}
\!\left(
\frac{\partial \boldsymbol{\sigma}^{i}}
     {\partial \boldsymbol{\varepsilon}^{i}}
\right)
\\
\displaystyle
\sum_{i=0}^{2^N-1}
v_f^i (\mathbf O^i)^{\mathsf T}
\frac{\partial \mathbf D^i}
     {\partial \boldsymbol{\varepsilon}^{i}}
\end{bmatrix},
\\
\label{eq:sensitivity_E_compact_inverse}
\begin{bmatrix}
\dfrac{\partial \mathbf A}{\partial \bar{\mathbf E}}
\\
\dfrac{\partial \mathbf B}{\partial \bar{\mathbf E}}
\end{bmatrix}
&=
-
\mathbf J^{-1}
\begin{bmatrix}
\displaystyle
\sum_{i=0}^{2^N-1}
v_f^i (\mathbf M^i)^{\mathsf T}
\frac{\partial \,\mathrm{vec}^{\boldsymbol{\sigma}}(\boldsymbol{\sigma}^{i})}
     {\partial \mathbf E^{i}}
\\[14pt]
\displaystyle
\sum_{i=0}^{2^N-1}
v_f^i (\mathbf O^i)^{\mathsf T}
\frac{\partial \mathbf D^i}
     {\partial \mathbf E^{i}}
\end{bmatrix}.
\end{align}

\subsection{Offline training}

The offline training stage identifies the hierarchical topological parameters of the proposed PDMN from homogenized linear electroelastic stiffness data. These parameters define the reduced microstructural representation of the network and provide the basis for the subsequent nonlinear online prediction. Specifically, the trainable parameter set is given by
\begin{equation}
\mathcal{F}
=
\left\{
\phi^{d}_{p},\,
\theta^{d}_{p},\,
W^{i}
\right\},
\qquad
i=0,\dots,2^N-1,\;
d=0,\dots,N-1,\;
p=0,\dots,2^d-1.
\end{equation}

These parameters are optimized so that the homogenized electroelastic stiffness predicted by the PDMN matches the corresponding DNS reference response. For a given parameter realization, the effective stiffness is computed recursively in a bottom-up traversal of the binary tree through the two-phase homogenization operator $\mathbb{H}_2(\cdot)$ (the generalized binary operator $\widehat{\mathbb{H}}_2$ of Eq.~\eqref{eq:H2_function}, with each child's volume fraction obtained from the node weights $W^i$ and hence omitted from the argument list):

\begin{equation}
\mathbf{C}^{d}_{p}
=
\begin{cases}
\displaystyle
\mathbb{H}_2
\bigl(
\mathbf{C}^{d+1}_{2p},
\mathbf{C}^{d+1}_{2p+1},
\mathbf{N}^{d}_{p}
\bigr),
& d < N-1, \\
\displaystyle
\mathbb{H}_2
\bigl(
\mathbf{C}^{p_1},
\mathbf{C}^{p_2},
\mathbf{N}^{d}_{p}
\bigr),
& d = N-1,
\end{cases}
\end{equation}
where $\mathbf{C}^{p_1}$ and $\mathbf{C}^{p_2}$ are the electroelastic stiffness matrices of the two constituent phases (phases~1 and~2), assigned to the terminal material nodes by the index parity introduced above.
from which the effective electroelastic stiffness of the network is obtained as
\begin{equation}
\bar{\mathbf{C}}^{\mathrm{PDMN}}=\mathbf{C}^{0}_{0}.
\end{equation}

The trainable parameters are determined by minimizing the relative
Frobenius-norm error between the PDMN prediction and the DNS reference over a
mini-batch of samples, namely
\begin{equation}
\label{eq:loss_function_revised}
\mathcal{L}(\mathcal{F})
=
\frac{1}{N_{\mathrm{batch}}}
\sum_{i=1}^{N_{\mathrm{batch}}}
\frac{
\left\|
\bar{\mathbf{C}}^{\mathrm{DNS}}_i
-
\bar{\mathbf{C}}^{\mathrm{PDMN}}_i(\mathcal{F})
\right\|_F^2
}{
\left\|
\bar{\mathbf{C}}^{\mathrm{DNS}}_i
\right\|_F^2
}.
\end{equation}
Here, $N_{\mathrm{batch}}$ denotes the mini-batch size, while
$\bar{\mathbf{C}}^{\mathrm{DNS}}_i$ and
$\bar{\mathbf{C}}^{\mathrm{PDMN}}_i$ represent the DNS-based and
PDMN-predicted homogenized electroelastic stiffness matrices of the $i$th
sample, respectively.

For each constituent phase, the linear electroelastic constitutive behavior is
represented in the $e$-form as
\begin{equation}
\label{eq:electroelastic_strain_charge}
\begin{bmatrix}
\sigma_{11}\\
\sigma_{22}\\
\sigma_{33}\\
\sigma_{23}\\
\sigma_{13}\\
\sigma_{12}\\
\hdashline
D_1\\
D_2\\
D_3
\end{bmatrix}
=
\left[
\begin{array}{cccccc:ccc}
C_{11}^E & C_{12}^E & C_{13}^E & 0 & 0 & 0 & 0 & 0 & -e_{31}\\
C_{12}^E & C_{11}^E & C_{13}^E & 0 & 0 & 0 & 0 & 0 & -e_{31}\\
C_{13}^E & C_{13}^E & C_{33}^E & 0 & 0 & 0 & 0 & 0 & -e_{33}\\
0 & 0 & 0 & C_{44}^E & 0 & 0 & 0 & -e_{15} & 0\\
0 & 0 & 0 & 0 & C_{44}^E & 0 & -e_{15} & 0 & 0\\
0 & 0 & 0 & 0 & 0 & \dfrac{C_{11}^E - C_{12}^E}{2} & 0 & 0 & 0\\
\hdashline
0 & 0 & 0 & 0 & e_{15} & 0 & \kappa_{11}^{\varepsilon} & 0 & 0\\
0 & 0 & 0 & e_{15} & 0 & 0 & 0 & \kappa_{11}^{\varepsilon} & 0\\
e_{31} & e_{31} & e_{33} & 0 & 0 & 0 & 0 & 0 & \kappa_{33}^{\varepsilon}
\end{array}
\right]
\begin{bmatrix}
\varepsilon_{11}\\
\varepsilon_{22}\\
\varepsilon_{33}\\
2\varepsilon_{23}\\
2\varepsilon_{13}\\
2\varepsilon_{12}\\
\hdashline
E_1\\
E_2\\
E_3
\end{bmatrix}.
\end{equation}

Accordingly, each phase is parameterized by the transversely isotropic material parameter vector
\begin{equation}
\label{eq:theta_phase}
\boldsymbol{\theta}^{(m)}
=
\left\{
C_{11,m}^{E},\,
C_{12,m}^{E},\,
C_{13,m}^{E},\,
C_{33,m}^{E},\,
C_{44,m}^{E},\,
e_{31,m},\,
e_{33,m},\,
e_{15,m},\,
\kappa_{11,m}^{\varepsilon},\,
\kappa_{33,m}^{\varepsilon}
\right\},
\qquad m\in\{p_1,p_2\}.
\end{equation}

The offline dataset consists of 500 triplets,
\(
\{
\mathbf{C}^{p_1},
\mathbf{C}^{p_2},
\bar{\mathbf{C}}^{\mathrm{DNS}}
\}
\),
of which 400 are used for training and the remaining 100 for validation. Here, $\mathbf{C}^{p_1}$ and $\mathbf{C}^{p_2}$ denote the phase-wise electroelastic constitutive matrices, and $\bar{\mathbf{C}}^{\mathrm{DNS}}$ is the corresponding homogenized response computed by DNS. The reference homogenized stiffness matrices are obtained from an ABAQUS user-defined element (UEL) implementation with periodic boundary conditions.

The objective of the offline dataset is to expose the PDMN to the full range of two-phase electroelastic stiffness combinations that it may encounter during online prediction. Because the network topology is identified once and reused for arbitrary constituents, the training data must densely cover the elastic, piezoelectric, and dielectric properties of each phase, as well as the contrast between the two phases, rather than a handful of specific materials. The constituent properties are therefore generated by a randomized reject-sampling procedure, summarized in Algorithm~\ref{alg:reject_sampling_piezo}, that spans a wide range of magnitudes and phase contrasts while discarding physically inadmissible combinations. In this way, the topology learned offline generalizes to the constituent combinations assigned at the online stage, which are not known a priori during training.

Each phase is first assigned a normalized, transversely isotropic base parameter set. For phase $m\in\{p_1,p_2\}$, the elastic constants are sampled as
\begin{equation}
\label{eq:base_elastic_sampling}
\begin{aligned}
C_{11,m}^{E}&=1, \\
C_{12,m}^{E},\,C_{13,m}^{E}&\sim U(0.1,0.7), \\
C_{33,m}^{E}&\sim U(0.5,2.0), \\
C_{44,m}^{E}&\sim U(0.1,0.5),
\end{aligned}
\end{equation}
where $U(a,b)$ denotes the uniform distribution on $[a,b]$, and $C_{11,m}^{E}$ is fixed to unity to remove the overall elastic scale at this stage. The piezoelectric and dielectric coefficients are sampled as
\begin{equation}
\label{eq:base_piezo_sampling}
\begin{aligned}
e_{31,m}&=s_{31}, \\
e_{33,m}&\sim U(0.8,8.0), \\
e_{15,m}&=s_{15}\,r_{15},\quad r_{15}\sim U(0.5,20.0), \\
\kappa_{11,m}^{\varepsilon},\,\kappa_{33,m}^{\varepsilon}&\sim U(0.7,2.0),
\end{aligned}
\end{equation}
where the signs $s_{31},s_{15}\sim U\{-1,+1\}$ are drawn independently so that both polarities of the transverse and shear piezoelectric coupling are represented. Drawing one such base set is summarized in Algorithm~\ref{alg:sample_one_phase_base}.

The two base sets are then rescaled to physical magnitudes with a controlled inter-phase contrast. The contrast of each property group is drawn on a logarithmic scale,
\begin{equation}
\label{eq:contrast_sampling}
\ell_{C},\;\ell_{e},\;\ell_{\kappa}\sim U(-3,3),
\end{equation}
and target magnitudes are assigned to the two phases, relative to the reference scales $C_{\mathrm{ref}}=10^{9}\,\mathrm{Pa}$, $e_{\mathrm{ref}}=1.0\,\mathrm{C/m^{2}}$, and $\kappa_{\mathrm{ref}}=10^{-9}\,\mathrm{F/m}$, as
\begin{equation}
\label{eq:contrast_targets}
\begin{aligned}
\bar{C}_{p_1}&=C_{\mathrm{ref}}, & \bar{C}_{p_2}&=C_{\mathrm{ref}}\,10^{\ell_{C}}, \\
\bar{e}_{p_1}&=e_{\mathrm{ref}}, & \bar{e}_{p_2}&=e_{\mathrm{ref}}\,10^{\ell_{e}}, \\
\bar{\kappa}_{p_1}&=\kappa_{\mathrm{ref}}, & \bar{\kappa}_{p_2}&=\kappa_{\mathrm{ref}}\,10^{\ell_{\kappa}},
\end{aligned}
\end{equation}
so that phase $p_1$ is held at the reference scale while phase $p_2$ carries the sampled contrast. The mean magnitude of each property group of phase $m$ is measured by
\begin{equation}
\label{eq:group_scales}
\begin{aligned}
C_{\mathrm{scale},m}&=\bigl(C_{11,m}^{E}C_{33,m}^{E}C_{44,m}^{E}\bigr)^{1/3}, \\
e_{\mathrm{scale},m}&=\sqrt{\tfrac{1}{3}\bigl(e_{31,m}^{2}+e_{33,m}^{2}+e_{15,m}^{2}\bigr)}, \\
\kappa_{\mathrm{scale},m}&=\bigl(\kappa_{11,m}^{\varepsilon}\kappa_{33,m}^{\varepsilon}\bigr)^{1/2},
\end{aligned}
\end{equation}
and every elastic, piezoelectric, and dielectric component of phase $m$ is rescaled by the corresponding factor,
\begin{equation}
\label{eq:rescaling}
\begin{aligned}
C_{ab,m}^{E}&\leftarrow \frac{\bar{C}_{m}}{C_{\mathrm{scale},m}}\,C_{ab,m}^{E}, \\
e_{ab,m}&\leftarrow \frac{\bar{e}_{m}}{e_{\mathrm{scale},m}}\,e_{ab,m}, \\
\kappa_{ab,m}^{\varepsilon}&\leftarrow \frac{\bar{\kappa}_{m}}{\kappa_{\mathrm{scale},m}}\,\kappa_{ab,m}^{\varepsilon}.
\end{aligned}
\end{equation}
Holding phase $p_1$ at the reference scale removes the redundant overall scaling of the RVE, while the logarithmic contrasts in Eq.~\eqref{eq:contrast_sampling} allow the two phases to differ by up to three orders of magnitude in each property group.

Finally, each candidate pair is accepted only if both phases satisfy the admissibility conditions for a physically meaningful linear electroelastic constitutive response, in particular, positive-definite elastic and dielectric moduli; inadmissible draws are rejected and resampled until $N_{\mathrm{samples}}$ admissible pairs are obtained.

\begin{algorithm}[htbp]
\caption{Reject sampling of two-phase transversely isotropic piezoelectric material parameters (base sampling: Eqs.~\eqref{eq:base_elastic_sampling}--\eqref{eq:base_piezo_sampling}; contrast and rescaling: Eqs.~\eqref{eq:contrast_sampling}--\eqref{eq:rescaling}).}
\label{alg:reject_sampling_piezo}
\begin{algorithmic}[1]
\Require target number of samples $N_{\mathrm{samples}}$, maximum number of trials $N_{\max}$
\Ensure dataset $\mathcal{D}$ of admissible two-phase parameter pairs
\State $\mathcal{D} \leftarrow [\,]$,\quad $n_{\mathrm{trial}} \leftarrow 0$
\While{$|\mathcal{D}| < N_{\mathrm{samples}}$ \textbf{and} $n_{\mathrm{trial}} < N_{\max}$}
    \State $n_{\mathrm{trial}} \leftarrow n_{\mathrm{trial}} + 1$
    \State $\boldsymbol{\theta}^{p_1} \leftarrow \textsc{SampleOnePhaseBase}()$
    \State $\boldsymbol{\theta}^{p_2} \leftarrow \textsc{SampleOnePhaseBase}()$
    \State $(\boldsymbol{\theta}^{p_1}, \boldsymbol{\theta}^{p_2}) \leftarrow \textsc{ApplyPhaseContrast}(\boldsymbol{\theta}^{p_1}, \boldsymbol{\theta}^{p_2})$
    \If{$\textsc{IsAdmissible}(\boldsymbol{\theta}^{p_1})$ \textbf{and} $\textsc{IsAdmissible}(\boldsymbol{\theta}^{p_2})$}
        \State append $(\boldsymbol{\theta}^{p_1}, \boldsymbol{\theta}^{p_2})$ to $\mathcal{D}$
    \EndIf
\EndWhile
\State \Return $\mathcal{D}$
\end{algorithmic}
\end{algorithm}

\begin{algorithm}[htbp]
\caption{\textsc{SampleOnePhaseBase}: draw one normalized base parameter set}
\label{alg:sample_one_phase_base}
\begin{algorithmic}[1]
\Ensure normalized transversely isotropic parameter set $\boldsymbol{\theta}$
\State $C_{11}^{E} \leftarrow 1$
\State sample $C_{12}^{E}, C_{13}^{E}, C_{33}^{E}, C_{44}^{E}$ from the elastic ranges in Eq.~\eqref{eq:base_elastic_sampling}
\State sample signs $s_{31}, s_{15} \sim U\{-1,+1\}$
\State $e_{31} \leftarrow s_{31}$;\quad $e_{33} \sim U(0.8,8.0)$
\State $e_{15} \leftarrow s_{15}\,r_{15}$ with $r_{15}\sim U(0.5,20.0)$
\State sample $\kappa_{11}^{\varepsilon}, \kappa_{33}^{\varepsilon} \sim U(0.7,2.0)$
\State \Return $\boldsymbol{\theta} = \{C_{11}^{E}, C_{12}^{E}, C_{13}^{E}, C_{33}^{E}, C_{44}^{E}, e_{31}, e_{33}, e_{15}, \kappa_{11}^{\varepsilon}, \kappa_{33}^{\varepsilon}\}$
\end{algorithmic}
\end{algorithm}

\newpage

\subsection{Online prediction}
\label{subsec:online_prediction}

During online prediction, the trained PDMN acts as a reduced-order surrogate model to efficiently evaluate the nonlinear coupled electromechanical response of piezoelectric materials. In contrast to the offline training stage, where the network parameters are identified, the online stage focuses on determining the constitutive response under prescribed macroscopic loading conditions using an iterative, fully coupled solution scheme.

In the PDMN formulation, the mechanical and electric fields are intrinsically coupled through the interaction variables associated with the hierarchical network architecture. As a result, the macroscopic response cannot be obtained from a single forward evaluation; instead, an iterative solution procedure is required to simultaneously enforce mechanical equilibrium and electric flux continuity across all interaction interfaces.

The overall online prediction procedure of the PDMN framework is summarized in Algorithm~\ref{alg:pdmn_online_prediction}. The algorithm begins by initializing the interaction variables, which define the initial state of the microscale field fluctuations. At each iteration, the macroscopic strain and electric field are downscaled to the material nodes, where the local constitutive laws are evaluated to obtain the stress, the electric displacement, and their corresponding tangent operators. From these local responses, the global residuals associated with mechanical equilibrium and electric continuity are assembled and iteratively driven to zero.

Upon convergence, an upscaling step is performed to compute the homogenized stress, homogenized electric displacement, and the consistent macroscopic tangent operators. This hierarchical procedure ensures consistency between microscale electromechanical interactions and the macroscopic constitutive response, enabling accurate prediction of anisotropic and nonlinear electromechanical behavior in piezoelectric materials.

\begin{algorithm}[H]
\caption{Online prediction algorithm of the PDMN at a loading increment}
\label{alg:pdmn_online_prediction}
\begin{algorithmic}[1]

\Statex \textbf{Input:}
$\bar{\boldsymbol{\varepsilon}},\,
\bar{\mathbf E},\,
t,\,
\Delta t$
\Statex \textit{(here $t$ denotes the current time and $\Delta t$ the time increment, required by history-dependent constituents)}

\Statex \textbf{Parameters:}
$\mathrm{tol}_{\mathrm{mech}}^{\mathrm{abs}},\,
\mathrm{tol}_{\mathrm{elec}}^{\mathrm{abs}},\,
\mathrm{tol}_{\mathrm{rel}},\,
\mathrm{iter}_{\max}$

\If{$t = 0$}
    \State $\mathbf A \gets \mathbf 0,\qquad \mathbf B \gets \mathbf 0$
\EndIf

\State $k \gets 0$

\While{$k < \mathrm{iter}_{\max}$}
    \State $k \gets k + 1$

    \State $(\boldsymbol{\varepsilon}^i,\mathbf E^i)
    \gets
    \textsc{LocalizeFields}\!\left(
    \bar{\boldsymbol{\varepsilon}},
    \bar{\mathbf E},
    \mathbf A,
    \mathbf B
    \right)$
    \Comment{Eqs.~\eqref{eq:interaction_mapping_strain_vec}--\eqref{eq:interaction_mapping_electric_vec}}

    \State $\left(
    \boldsymbol{\sigma}^i,
    \mathbf D^i,
    \mathbf C_{\sigma\varepsilon}^i,
    \mathbf C_{\sigma E}^i,
    \mathbf C_{D\varepsilon}^i,
    \mathbf C_{DE}^i
    \right)
    \gets
    \textsc{EvaluateLocalResponse}\!\left(
    \boldsymbol{\varepsilon}^i,
    \mathbf E^i,
    \Delta t
    \right)$
    \Comment{Eqs.~\eqref{eq:localLaw1}--\eqref{eq:localLaw2}}

    \State $r_{\mathrm{mech}}, r_{\mathrm{elec}}
    \gets
    \textsc{ComputeResiduals}\!\left(
    \boldsymbol{\sigma}^i,
    \mathbf D^i
    \right)$
    \Comment{Eqs.~\eqref{eq:residual_mech}--\eqref{eq:residual_elec}}

    \If{$k = 1$}
        \State $r_{0,\mathrm{mech}} \gets \max\!\left(r_{\mathrm{mech}}, 10^{-16}\right)$
        \State $r_{0,\mathrm{elec}} \gets \max\!\left(r_{\mathrm{elec}}, 10^{-16}\right)$
    \EndIf

    \State $r_{\mathrm{rel,mech}} \gets r_{\mathrm{mech}} / r_{0,\mathrm{mech}}$
    \State $r_{\mathrm{rel,elec}} \gets r_{\mathrm{elec}} / r_{0,\mathrm{elec}}$

    \State $\mathrm{conv}_{\mathrm{mech}} \gets
    \left(r_{\mathrm{mech}} < \mathrm{tol}_{\mathrm{mech}}^{\mathrm{abs}}\right)
    \;\textbf{or}\;
    \left(r_{\mathrm{rel,mech}} < \mathrm{tol}_{\mathrm{rel}}\right)$

    \State $\mathrm{conv}_{\mathrm{elec}} \gets
    \left(r_{\mathrm{elec}} < \mathrm{tol}_{\mathrm{elec}}^{\mathrm{abs}}\right)
    \;\textbf{or}\;
    \left(r_{\mathrm{rel,elec}} < \mathrm{tol}_{\mathrm{rel}}\right)$

    \If{$\mathrm{conv}_{\mathrm{mech}} \;\textbf{and}\; \mathrm{conv}_{\mathrm{elec}}$}
        \State \textbf{break}
    \EndIf

    \State $\mathbf r \gets
    \begin{bmatrix}
    \mathbf r_{\mathrm{mech}}\\
    \mathbf r_{\mathrm{elec}}
    \end{bmatrix}$

    \State $\mathbf J
    \gets
    \textsc{AssembleJacobian}\!\left(
    \mathbf C_{\sigma\varepsilon}^i,
    \mathbf C_{\sigma E}^i,
    \mathbf C_{D\varepsilon}^i,
    \mathbf C_{DE}^i
    \right)$
    \Comment{Eqs.~\eqref{eq:coupled_newton}--\eqref{eq:drelec_dB}}

    \State Compute $\Delta \mathbf x$ from
    $\mathbf J\,\Delta \mathbf x = -\,\mathbf r$

    \State
    $\begin{bmatrix}
    \mathbf A\\
    \mathbf B
    \end{bmatrix}
    \gets
    \begin{bmatrix}
    \mathbf A\\
    \mathbf B
    \end{bmatrix}
    + \Delta \mathbf x$
\EndWhile

\Statex \textit{Macroscopic response and consistent tangent}

\State $(\bar{\boldsymbol{\sigma}},\bar{\mathbf D})
\gets
\textsc{ComputeResponse}\!\left(
\boldsymbol{\sigma}^i,
\mathbf D^i
\right)$
\Comment{Eq.~\eqref{eq:macro_sigma_avg}}

\State $\widehat{\mathbf C}^{\mathrm{alg}}_{\mathrm{hom}}
\gets
\textsc{ComputeTangent}\!\left(
\mathbf C_{\sigma\varepsilon}^i,
\mathbf C_{\sigma E}^i,
\mathbf C_{D\varepsilon}^i,
\mathbf C_{DE}^i
\right)$
\Comment{Eqs.~\eqref{eq:Chat_general}--\eqref{eq:sensitivity_E_compact_inverse}}

\Statex \textbf{Output:}
$\bar{\boldsymbol{\sigma}},\,
\bar{\mathbf D},\,
\widehat{\mathbf C}^{\mathrm{alg}}_{\mathrm{hom}}$

\end{algorithmic}
\end{algorithm}

\section{Results}

This section evaluates the proposed PDMN framework from three complementary perspectives. First, the offline training performance is examined to assess whether the PDMN can learn the effective electroelastic response and recover the key microstructural features of the reference RVE. Second, in the online stage, the trained PDMN predicts the nonlinear electroelastic response under prescribed macroscopic electric and mechanical loading paths, with the results compared against DNS. Finally, the computational efficiency of the framework is quantified to demonstrate its suitability as a surrogate for repeated multiscale homogenization.

\subsection{Offline training}

\begin{figure}[htbp]
\centering
\includegraphics[width=0.5\textwidth]{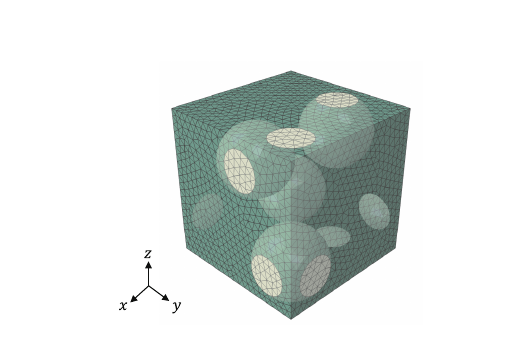}
\caption{
    Geometry of the RVE considered in this study.
} 
\label{fig:RVE_geometry}
\end{figure}

The RVE considered in this study is adopted from previous work~\cite{liu2019exploring, DYNA_website}. The discretized mesh contains 59,628 10-node tetrahedral (C3D10) elements, and the inclusion volume fraction is 22.6\%. The corresponding RVE geometry is shown in Fig.~\ref{fig:RVE_geometry}. This RVE is used throughout the offline training stage to generate the reference dataset for learning the homogenized electroelastic response.

\begin{figure}[htbp]
\centering
\includegraphics[width=1\textwidth]{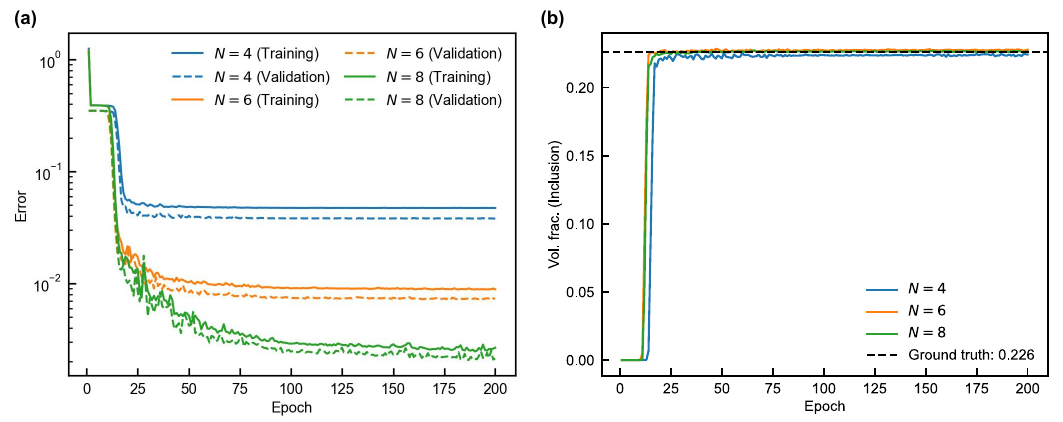}
\caption{(a) Training and validation curves and (b) predicted inclusion volume fraction for the PDMN at different hierarchical levels $N$.}

\label{fig:training_curve}
\end{figure}

Fig.~\ref{fig:training_curve}(a) summarizes the offline training results of the proposed PDMN for different hierarchical levels $N$. Both the training and validation errors decrease consistently as $N$ increases, indicating that a deeper hierarchy better represents the effective electroelastic response of the RVE. For the shallowest network, with $N=4$, the validation error remains at approximately 4--5\%, whereas for $N=8$ it falls below 0.3\%.

Further insight comes from the inclusion volume fraction reconstructed from the trained PDMN parameters. As shown in Fig.~\ref{fig:training_curve}(b) and summarized in Table~\ref{tab:offline_training_summary}, the predicted volume fraction approaches the reference value of 22.6\% as $N$ increases, reaching 0.2242 for $N=4$, 0.2278 for $N=6$, and 0.2267 for $N=8$. At the same time, a larger $N$ entails more trainable parameters and longer training time. These results show that increasing the hierarchical level improves both the prediction accuracy and the fidelity of the learned microstructural representation, albeit at a higher training cost.

\begin{table}[htbp]
\centering
\caption{Summary of offline training performance for PDMN models with different hierarchical levels $N$.}
\label{tab:offline_training_summary}
\begin{tabular}{cccc}
\hline
Hierarchical level $N$ & Number of parameters & Training time & Predicted volume fraction \\
\hline
4 & 46  & 56 s    & 0.2242 \\
6 & 190 & 4 min 58 s & 0.2278 \\
8 & 766 & 35 min 14 s & 0.2267 \\
\hline
\end{tabular}
\end{table}

\subsection{Online prediction of nonlinear electroelastic response}

\subsubsection{Constituent models and loading paths}

During the online prediction stage, the trained PDMN serves as a surrogate for coupled electromechanical homogenization. The PDMN parameters identified during offline training are held fixed, while the material nodes are assigned constituent-level constitutive models based on the phase arrangement in each case.

Three online prediction cases are investigated. Cases I and II examine the nonlinear electroelastic response of two-phase composites of polyvinylidene fluoride (PVDF) and lithium niobate (LiNbO$_3$) with reversed phase arrangements. In Case I, PVDF serves as the matrix phase and LiNbO$_3$ as the inclusion phase, whereas in Case II, LiNbO$_3$ serves as the matrix phase and PVDF as the inclusion phase. In both cases, PVDF is modeled as a linear electroelastic material, while LiNbO$_3$ is described by a nonlinear electroelastic constitutive law~\cite{tassi2023nonlinear}. For LiNbO$_3$, the stress and electric displacement are expressed as
\begin{equation}
\left\{
\begin{aligned}
\sigma_{ij}
&=
C_{ijlm}\,\varepsilon_{lm}
+\frac{1}{2}\tilde{C}_{ijlmpq}\,\varepsilon_{lm}\varepsilon_{pq}
- e_{ijn}E_n
-\frac{1}{2} \beta_{ijnr}E_nE_r
-\tilde{e}_{ijlmn}\,\varepsilon_{lm}E_n,
\\[4pt]
D_k
&=
e_{klm}\,\varepsilon_{lm}
+\frac{1}{2}\tilde{e}_{klmpq}\,\varepsilon_{lm}\varepsilon_{pq}
+\kappa^{\varepsilon}_{kn}E_n
+\frac{1}{2}\chi^{\varepsilon}_{knr}E_nE_r
+ \beta_{klmn}\,\varepsilon_{lm}E_n .
\end{aligned}
\right.
\label{eq:nonlinear_electroelastic_law}
\end{equation}
In Eq.~\eqref{eq:nonlinear_electroelastic_law}, all indices range over $\{1,2,3\}$ and repeated indices imply summation. The coefficients $C_{ijlm}$, $e_{ijn}$, and $\kappa^{\varepsilon}_{kn}$ are the full-tensor counterparts of the Voigt coefficients in Eq.~\eqref{eq:electroelastic_strain_charge}, whereas $\tilde{C}_{ijlmpq}$, $\chi^{\varepsilon}_{knr}$, $\beta_{ijnr}$, and $\tilde{e}_{ijlmn}$ are the third-order elastic, nonlinear dielectric, electrostrictive, and nonlinear electroelastic constants, respectively.
The linear electroelastic coefficients of PVDF and LiNbO$_3$, together with the nonlinear elastic, dielectric, electrostrictive, and electroelastic constants of LiNbO$_3$, are taken from~\cite{tassi2023nonlinear}. These material parameters are summarized in Tables~\ref{tab:linear_material_parameters} and~\ref{tab:nonlinear_parameters_LN}.

For Cases I and II, two loading paths are considered to probe the coupled electromechanical response. The first is electric-field loading along the $E_3$ direction under strain-free conditions, and the second is uniaxial mechanical loading in the $\varepsilon_{33}$ direction under electric-field-free conditions. For both paths, the PDMN predictions are compared with the corresponding DNS results for the homogenized stress $\sigma_{33}$ and the electric displacement $D_3$.

Case III evaluates the capability of the trained PDMN to predict time-dependent material responses in a coupled electromechanical setting. Here, the matrix phase is described by a linear viscoelastic constitutive model, whereas the inclusion phase is modeled as a linear piezoelastic material. This configuration enables the assessment of a composite system in which mechanical relaxation in the matrix interacts with the electromechanical response of the piezoelastic inclusions. For the viscoelastic matrix, the stress response is formulated through the hereditary integral
\begin{equation}
\boldsymbol{\sigma}(t)
=
\int_{0}^{t}
2G(t-s)
\frac{d\boldsymbol{\varepsilon}^{\mathrm{dev}}(s)}{ds}
\,ds
+
\mathbf I
\int_{0}^{t}
K(t-s)
\frac{d\varepsilon^{\mathrm{vol}}(s)}{ds}
\,ds,
\label{eq:linear_viscoelastic_stress}
\end{equation}
where $G(t)$ and $K(t)$ denote the shear and bulk relaxation moduli,
respectively, $\boldsymbol{\varepsilon}^{\mathrm{dev}}$ is the deviatoric strain
tensor, $\varepsilon^{\mathrm{vol}}$ is the volumetric strain, and $\mathbf I$
is the second-order identity tensor.

The relaxation behavior is represented by a time-domain Prony series. The shear relaxation modulus is written as
\begin{equation}
G(t)
=
G_0
\left[
1
-
\sum_{k=1}^{n}
g_k
\left(
1-\exp\left(-\frac{t}{\tau_k}\right)
\right)
\right],
\label{eq:prony_shear_modulus}
\end{equation}
and the bulk relaxation modulus is given by
\begin{equation}
K(t)
=
K_0
\left[
1
-
\sum_{k=1}^{n}
k_k
\left(
1-\exp\left(-\frac{t}{\tau_k}\right)
\right)
\right],
\label{eq:prony_bulk_modulus}
\end{equation}
where $G_0$ and $K_0$ are the instantaneous shear and bulk moduli, respectively,
$g_k$ and $k_k$ are the normalized Prony coefficients associated with shear and
bulk relaxation, and $\tau_k$ is the relaxation time of the $k$th Prony term.
The material parameters used for the matrix phase and the inclusion phase are summarized in
Tables~\ref{tab:viscoelastic_matrix_parameters}
and~\ref{tab:linear_piezoelectric_inclusion_parameters}, respectively.

\begin{table*}[htbp]
\centering
\caption{Linear electroelastic coefficients of LiNbO$_3$ and PVDF~\cite{tassi2023nonlinear}.}
\label{tab:linear_material_parameters}
\renewcommand{\arraystretch}{1.15}
\setlength{\tabcolsep}{6pt}

\begin{tabular}{lccccc}
\toprule
Material
& $C_{11}$ (GPa)
& $C_{12}$ (GPa)
& $C_{13}$ (GPa)
& $C_{33}$ (GPa)
& $C_{44}$ (GPa) \\
\midrule
LiNbO$_3$ & 203.0 & 53.0 & 75.0 & 245 & 60.0 \\
PVDF      & 2.26  & 1.07 & 1.07 & 2.26 & 0.775 \\
\bottomrule
\end{tabular}

\vspace{0.8em}

\begin{tabular}{lccccc}
\toprule
Material
& $e_{31}$ (C/m$^2$)
& $e_{33}$ (C/m$^2$)
& $e_{15}$ (C/m$^2$)
& $\kappa_{11}^{\varepsilon}$ ($10^{-9}$ F/m)
& $\kappa_{33}^{\varepsilon}$ ($10^{-9}$ F/m) \\
\midrule
LiNbO$_3$ & 0.2   & 1.3   & 3.7     & 0.389  & 0.257  \\
PVDF      & 0.046 & 0.046 & -0.0391 & 0.1062 & 0.1062 \\
\bottomrule
\end{tabular}
\end{table*}

\begin{table*}[htbp]
\centering
\caption{Nonlinear material parameters of LiNbO$_3$ used in the online prediction stage~\cite{tassi2023nonlinear}. Contracted (Voigt) indices are used in this table, with $4\!\rightarrow\!23$, $5\!\rightarrow\!13$, and $6\!\rightarrow\!12$.}
\label{tab:nonlinear_parameters_LN}
\renewcommand{\arraystretch}{1.15}
\setlength{\tabcolsep}{6pt}

\begin{tabular*}{0.98\textwidth}{@{\extracolsep{\fill}}cccccccc}
\toprule
\multicolumn{8}{c}{\textbf{Nonlinear elastic constants} $(\times 10^{11}\,\mathrm{N}/\mathrm{m}^2)$} \\
\midrule
$C_{111}=C_{222}$ & $C_{112}$ & $C_{113}$ & $C_{122}$ & $C_{123}$ & $C_{133}=C_{233}$ & $C_{333}$ & $C_{444}$ \\
\midrule
$-21.2$ & $-5.3$ & $-5.7$ & $-3.2$ & $-2.5$ & $-7.8$ & $-29.6$ & $-0.3$ \\
\bottomrule
\end{tabular*}

\vspace{0.8em}

\begin{tabular}{ccc}
\toprule
\multicolumn{3}{c}{\textbf{Nonlinear dielectric constants} $(\times 10^{-19}\,\mathrm{F}/\mathrm{V})$} \\
\midrule
$\chi^{\varepsilon}_{111}$ & $\chi^{\varepsilon}_{222}$ & $\chi^{\varepsilon}_{333}$ \\
\midrule
$-2.81$ & $-2.40$ & $-2.91$ \\
\bottomrule
\end{tabular}

\vspace{0.8em}

\begin{tabular*}{0.72\textwidth}{@{\extracolsep{\fill}}cccccc}
\toprule
\multicolumn{6}{c}{\textbf{Electrostriction coefficients} $(\times 10^{-9}\,\mathrm{F}/\mathrm{m})$} \\
\midrule
$b_{111}=b_{222}$ & $b_{112}$ & $b_{133}$ & $b_{333}$ & $b_{423}$ & $b_{612}$ \\
\midrule
$1.11$ & $2.19$ & $2.32$ & $-2.76$ & $-1.83$ & $-0.54$ \\
\bottomrule
\end{tabular*}

\vspace{0.8em}

\begin{tabular*}{1.1\textwidth}{@{\extracolsep{\fill}}ccccccccccc}
\toprule
\multicolumn{11}{c}{\textbf{Electroelasticity constants} $(\mathrm{C}/\mathrm{m}^2)$} \\
\midrule
$\tilde{e}_{111}$ & $\tilde{e}_{121}$ & $\tilde{e}_{131}$ & $\tilde{e}_{122}$ & $\tilde{e}_{132}$ & $\tilde{e}_{133}$
& $\tilde{e}_{441}$ & $\tilde{e}_{232}$ & $\tilde{e}_{333}$ & $\tilde{e}_{443}$ & $\tilde{e}_{633}$ \\
\midrule
$15.0$ & $-26.2$ & $14.7$ & $-10.3$ & $-0.9$ & $-10.0$
& $20.3$ & $0.9$ & $-17.3$ & $-10.5$ & $0.85$ \\
\bottomrule
\end{tabular*}

\end{table*}

\begin{table}[htbp]
\centering
\caption{Viscoelastic material parameters of the matrix phase used in Case III.}
\label{tab:viscoelastic_matrix_parameters}
\renewcommand{\arraystretch}{1.15}
\setlength{\tabcolsep}{6pt}

\begin{tabular}{lc}
\toprule
Parameter & Value \\
\midrule
$G_0$ (GPa) & 1.70 \\
$K_0$ (GPa) & 6.01 \\
\bottomrule
\end{tabular}

\vspace{0.8em}

\begin{tabular}{lccc}
\toprule
Prony term & 1 & 2 & 3 \\
\midrule
$\tau_k$ (s) & 9.6 & 372 & 9887 \\
$g_k$ (-) & 0.03807 & 0.0458 & 0.0668 \\
$k_k$ (-) & 0 & 0 & 0 \\
\bottomrule
\end{tabular}
\end{table}

\begin{table}[htbp]
\centering
\caption{Linear piezoelectric material parameters of the inclusion phase used in Case III.}
\label{tab:linear_piezoelectric_inclusion_parameters}
\renewcommand{\arraystretch}{1.15}
\setlength{\tabcolsep}{6pt}

\begin{tabular}{lcccccc}
\toprule
Material
& $C_{11}$ (GPa)
& $C_{12}$ (GPa)
& $C_{13}$ (GPa)
& $C_{33}$ (GPa)
& $C_{44}$ (GPa) \\
\midrule
Inclusion
& 150.4
& 65.63
& 65.94
& 145.5
& 43.86 \\
\bottomrule
\end{tabular}

\vspace{0.8em}

\begin{tabular}{lccccc}
\toprule
Material
& $e_{31}$ (C/m$^2$)
& $e_{33}$ (C/m$^2$)
& $e_{15}$ (C/m$^2$)
& $\kappa_{11}^{\varepsilon}$ ($10^{-9}$ F/m)
& $\kappa_{33}^{\varepsilon}$ ($10^{-9}$ F/m) \\
\midrule
Inclusion
& -4.32
& 17.4
& 11.4
& 12.8
& 12.8 \\
\bottomrule
\end{tabular}
\end{table}

\newpage

\subsubsection{Case I: PVDF matrix with LiNbO$_3$ inclusion}

\begin{figure}[htbp]
    \centering
    \includegraphics[width=1.0\linewidth]{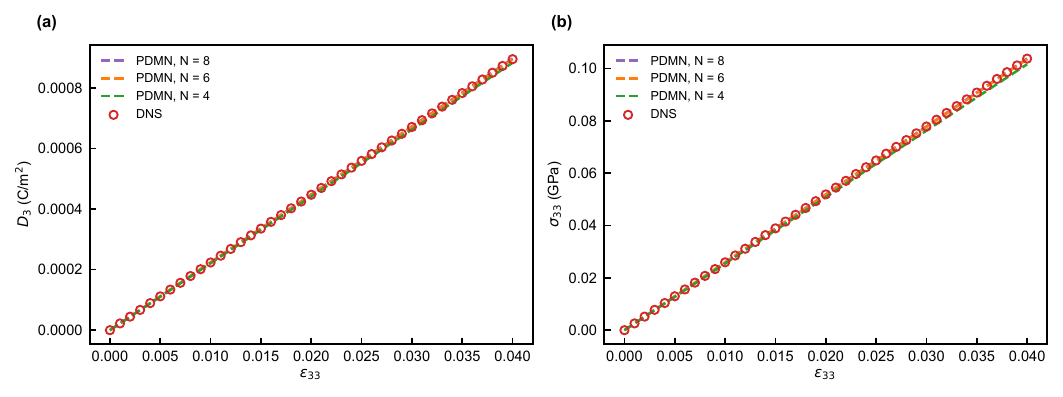}
    \caption{Effective electroelastic response of the Case I composite under
    $\varepsilon_{33}$-loading: 
    (a) $D_3$--$\varepsilon_{33}$ response and 
    (b) $\sigma_{33}$--$\varepsilon_{33}$ response.}
    \label{fig:PVDFLiNbO3_e33}
\end{figure}

\begin{figure}[htbp]
    \centering
    \includegraphics[width=1.0\linewidth]{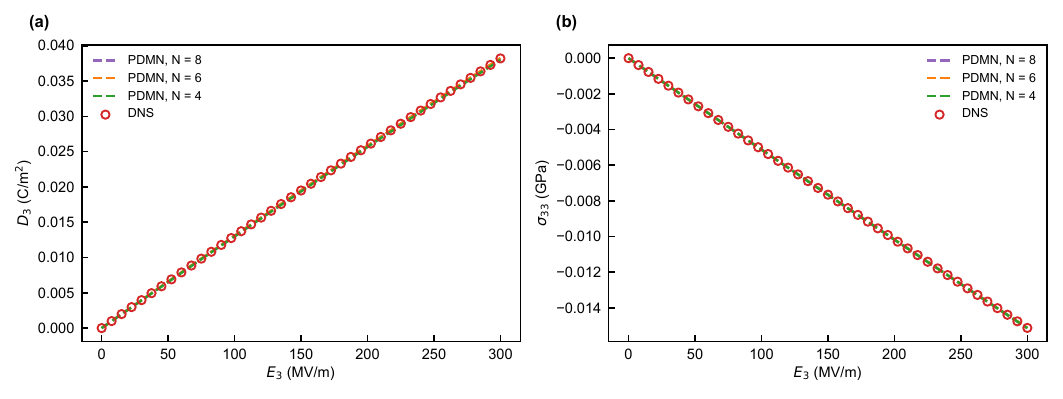}
    \caption{Effective electroelastic response of the Case I composite under
    $E_3$-loading: 
    (a) $D_3$--$E_3$ response and 
    (b) $\sigma_{33}$--$E_3$ response.}
    \label{fig:PVDFLiNbO3_E3}
\end{figure}

In Case I, PVDF serves as the matrix phase, while LiNbO$_3$ is embedded as the inclusion phase. The predicted effective $\sigma_{33}$ and $D_3$ responses under $\varepsilon_{33}$-loading are presented in Fig.~\ref{fig:PVDFLiNbO3_e33}, and the corresponding responses under $E_3$-loading are shown in Fig.~\ref{fig:PVDFLiNbO3_E3}.

For this phase arrangement, the overall electroelastic response is primarily
governed by the PVDF matrix. Consequently, the nonlinear contribution associated
with the LiNbO$_3$ inclusion is relatively limited at the macroscopic level.
The close agreement between the PDMN and DNS results for both mechanical and
electrical loading paths indicates that the proposed framework can accurately
predict homogenized electroelastic responses.

To quantitatively assess prediction accuracy, the mean relative error (MRE) and maximum relative error (MaxRE) are used, following established practice~\cite{huang2022microstructure}. These metrics provide normalized measures of the deviation from the DNS results and are defined as
\begin{equation}
\mathrm{MRE} =
\frac{
\frac{1}{n}\sum_{i=1}^{n}
\left| q_i^{\mathrm{DNS}} - q_i^{\mathrm{PDMN}} \right|
}{
\max_{i=1,\ldots,n} \left| q_i^{\mathrm{DNS}} \right|
},
\label{eq:mean_rel_error}
\end{equation}
and
\begin{equation}
\mathrm{MaxRE} =
\frac{
\max_{i=1,\ldots,n}
\left| q_i^{\mathrm{DNS}} - q_i^{\mathrm{PDMN}} \right|
}{
\max_{i=1,\ldots,n} \left| q_i^{\mathrm{DNS}} \right|
}.
\label{eq:max_rel_error}
\end{equation}
Here, $q_i$ denotes the scalar response quantity evaluated at the $i$th loading increment, such as $\sigma_{33}$ or $D_3$, and $n$ is the total number of loading increments.

The resulting metrics are summarized in Table~\ref{tab:PVDFLiNbO3_relative_error_metrics}. The overall trend is that the prediction error decreases as the hierarchical level $N$ increases. The agreement with DNS is excellent across all loading paths and response quantities; over the entire table, the largest deviation occurs for the $\sigma_{33}$ response under $\varepsilon_{33}$-loading, where even the shallowest network ($N=4$) attains a MaxRE of only $2.06\%$. For this loading path, increasing the network depth markedly improves the accuracy, with the $\sigma_{33}$ MaxRE reduced to $0.05\%$ and the MRE from $1.01\%$ to $0.03\%$ at $N=8$.

\begin{table}[htbp]
\centering
\caption{Mean-relative and max-relative error metrics of the PDMN predictions
for the Case I composite with a PVDF matrix and LiNbO$_3$ inclusion.}
\label{tab:PVDFLiNbO3_relative_error_metrics}
\renewcommand{\arraystretch}{1.15}
\setlength{\tabcolsep}{6pt}
\begin{tabular}{llccc}
\toprule
Loading path & Response & Model & MRE & MaxRE \\
\midrule
\multirow{6}{*}{$E_3$}
& \multirow{3}{*}{$D_3$}
& $N=8$ & $1.170965\times10^{-4}$ & $1.936738\times10^{-4}$ \\
& & $N=6$ & $4.492084\times10^{-4}$ & $8.525918\times10^{-4}$ \\
& & $N=4$ & $3.205039\times10^{-3}$ & $5.182282\times10^{-3}$ \\
\cmidrule(lr){2-5}
& \multirow{3}{*}{$\sigma_{33}$}
& $N=8$ & $3.671081\times10^{-4}$ & $6.787194\times10^{-4}$ \\
& & $N=6$ & $1.202995\times10^{-4}$ & $2.304811\times10^{-4}$ \\
& & $N=4$ & $3.682573\times10^{-4}$ & $6.965726\times10^{-4}$ \\
\midrule
\multirow{6}{*}{$\varepsilon_{33}$}
& \multirow{3}{*}{$D_3$}
& $N=8$ & $1.322730\times10^{-3}$ & $2.550655\times10^{-3}$ \\
& & $N=6$ & $2.149905\times10^{-4}$ & $5.777040\times10^{-4}$ \\
& & $N=4$ & $6.330506\times10^{-3}$ & $1.309928\times10^{-2}$ \\
\cmidrule(lr){2-5}
& \multirow{3}{*}{$\sigma_{33}$}
& $N=8$ & $2.774193\times10^{-4}$ & $4.968797\times10^{-4}$ \\
& & $N=6$ & $1.167757\times10^{-3}$ & $2.446633\times10^{-3}$ \\
& & $N=4$ & $1.006092\times10^{-2}$ & $2.064667\times10^{-2}$ \\
\bottomrule
\end{tabular}
\end{table}

\subsubsection{Case II: LiNbO$_3$ matrix with PVDF inclusion}

\begin{figure}[htbp]
    \centering
    \includegraphics[width=1.0\linewidth]{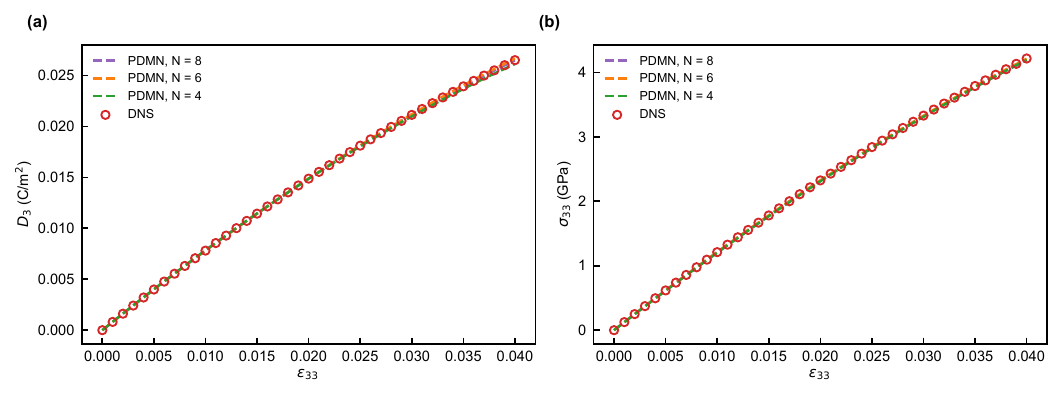}
    \caption{Effective electroelastic response of the Case II composite under
    $\varepsilon_{33}$-loading:
    (a) $D_3$--$\varepsilon_{33}$ response and
    (b) $\sigma_{33}$--$\varepsilon_{33}$ response.}
    \label{fig:LiNbO3PVDF_e33}
\end{figure}

\begin{figure}[htbp]
    \centering
    \includegraphics[width=1.0\linewidth]{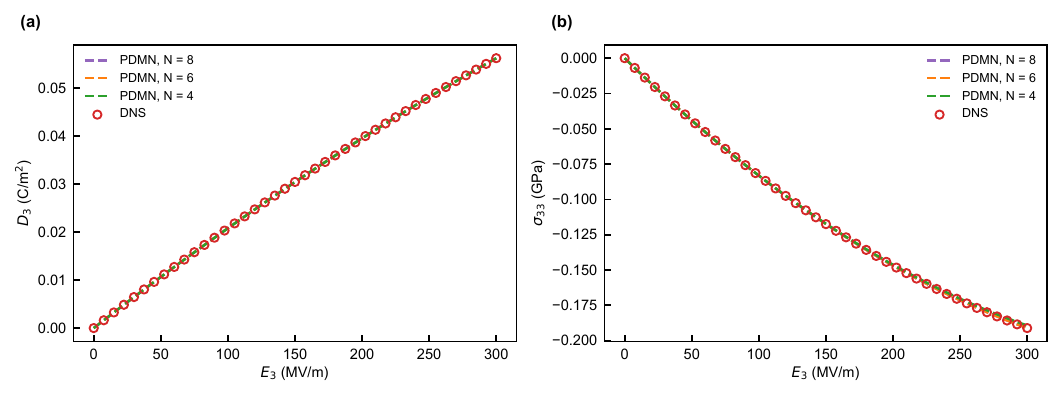}
    \caption{Effective electroelastic response of the Case II composite under
    $E_3$-loading:
    (a) $D_3$--$E_3$ response and
    (b) $\sigma_{33}$--$E_3$ response.}
    \label{fig:LiNbO3PVDF_E3}
\end{figure}

In Case II, LiNbO$_3$ serves as the matrix phase, while PVDF is embedded as the inclusion phase. The predicted effective $\sigma_{33}$ and $D_3$ responses under $\varepsilon_{33}$-loading are presented in Fig.~\ref{fig:LiNbO3PVDF_e33}, and the corresponding responses under $E_3$-loading are shown in Fig.~\ref{fig:LiNbO3PVDF_E3}.

In contrast to Case I, the effective response in this phase arrangement is primarily governed by the nonlinear LiNbO$_3$ matrix. As a result, the macroscopic electroelastic response exhibits more pronounced nonlinear characteristics. The close agreement between the PDMN and DNS results for both mechanical and electrical loading paths further confirms the capability of the proposed framework to predict homogenized nonlinear electroelastic responses.

The corresponding quantitative error metrics are summarized in Table~\ref{tab:LiNbO3PVDF_relative_error_metrics}. For all loading paths and response quantities, the PDMN predictions remain in close agreement with the DNS results. The errors generally decrease as network depth increases, although minor non-monotonic variations are observed in some responses due to the different sensitivities of the homogenized fields to the learned microstructural representation. In particular, the largest deviation over the entire table occurs for the $D_3$ response under $\varepsilon_{33}$-loading, where the shallowest network ($N=4$) gives a MaxRE of $1.41\%$; for this loading path, the error is markedly reduced as the network deepens, with the $D_3$ MaxRE falling to $0.08\%$ and the MRE from $0.36\%$ to $0.02\%$ at $N=8$.

\begin{table}[htbp]
\centering
\caption{Mean-relative and max-relative error metrics of the PDMN predictions
for the Case II composite with a LiNbO$_3$ matrix and PVDF inclusion.}
\label{tab:LiNbO3PVDF_relative_error_metrics}
\renewcommand{\arraystretch}{1.15}
\setlength{\tabcolsep}{6pt}
\begin{tabular}{llccc}
\toprule
Loading path & Response & Model & MRE & MaxRE \\
\midrule
\multirow{6}{*}{$E_3$}
& \multirow{3}{*}{$D_3$}
& $N=8$ & $3.440967\times10^{-5}$ & $5.850838\times10^{-5}$ \\
& & $N=6$ & $3.602411\times10^{-4}$ & $6.532641\times10^{-4}$ \\
& & $N=4$ & $3.874404\times10^{-4}$ & $5.698521\times10^{-4}$ \\
\cmidrule(lr){2-5}
& \multirow{3}{*}{$\sigma_{33}$}
& $N=8$ & $1.279438\times10^{-4}$ & $3.784392\times10^{-4}$ \\
& & $N=6$ & $3.546014\times10^{-4}$ & $5.520097\times10^{-4}$ \\
& & $N=4$ & $2.687744\times10^{-3}$ & $9.475319\times10^{-3}$ \\
\midrule
\multirow{6}{*}{$\varepsilon_{33}$}
& \multirow{3}{*}{$D_3$}
& $N=8$ & $2.223823\times10^{-4}$ & $7.705554\times10^{-4}$ \\
& & $N=6$ & $1.423888\times10^{-3}$ & $6.230075\times10^{-3}$ \\
& & $N=4$ & $3.595682\times10^{-3}$ & $1.412461\times10^{-2}$ \\
\cmidrule(lr){2-5}
& \multirow{3}{*}{$\sigma_{33}$}
& $N=8$ & $1.870710\times10^{-3}$ & $6.121412\times10^{-3}$ \\
& & $N=6$ & $3.513874\times10^{-4}$ & $7.996791\times10^{-4}$ \\
& & $N=4$ & $3.108762\times10^{-3}$ & $5.283474\times10^{-3}$ \\
\bottomrule
\end{tabular}
\end{table}

\subsubsection{Case III: Viscoelastic matrix with piezoelectric inclusion}

\begin{figure}[htbp]
    \centering
    \includegraphics[width=1.0\linewidth]{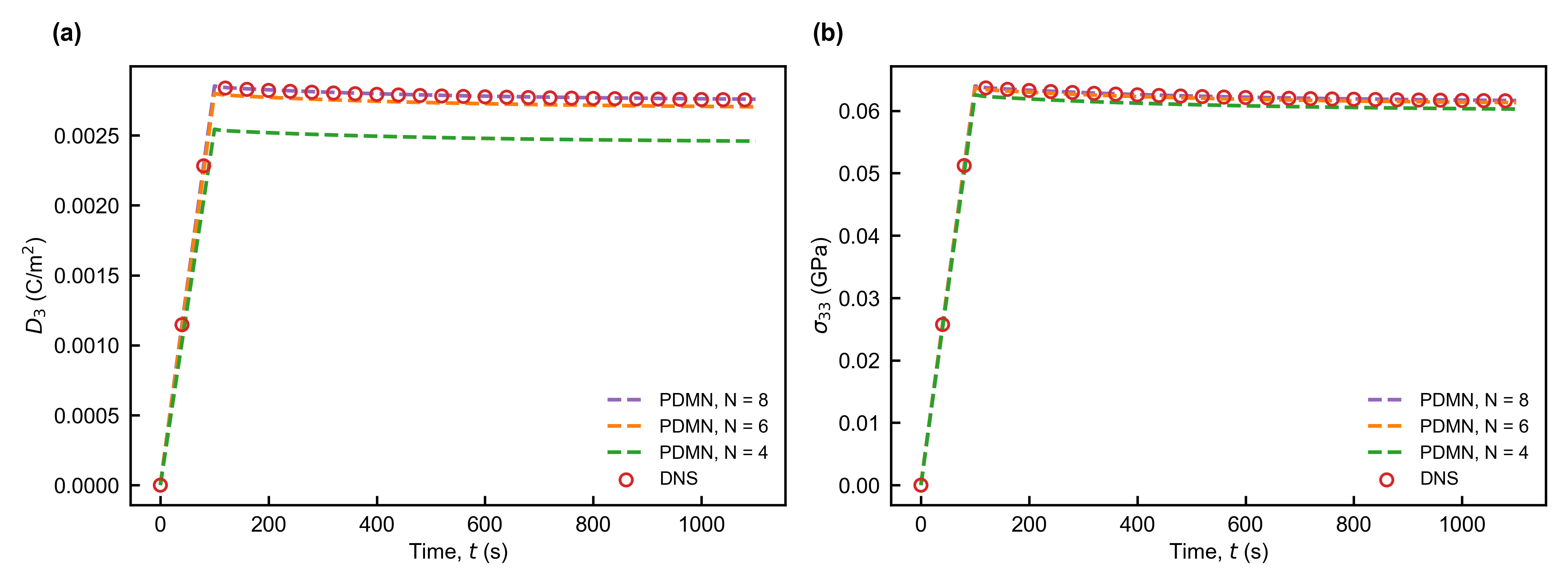}
    \caption{Effective electromechanical response of the Case III composite
    under $\varepsilon_{33}$-loading followed by a hold period:
    (a) temporal evolution of $\sigma_{33}$ and
    (b) temporal evolution of $D_3$.}
    \label{fig:composite_relax}
\end{figure}

Case III evaluates the capability of the PDMN to predict time-dependent coupled electromechanical responses. The composite consists of a linear viscoelastic matrix and a linear piezoelectric inclusion. The RVE is subjected to a ramp loading up to $\varepsilon_{33}=0.01$ over 100~s, followed by a 1000~s hold period to examine the stress-relaxation response under a prescribed macroscopic strain. The predicted effective $\sigma_{33}$ and $D_3$ responses are compared with DNS in Fig.~\ref{fig:composite_relax}, and the corresponding error metrics are summarized in Table~\ref{tab:case3_stress_relaxation_errors}.

During the hold period, the effective stress gradually decreases as the viscoelastic matrix relaxes. Although the inclusion is modeled as a linear piezoelectric material, the effective electric displacement $D_3$ likewise evolves with time. This behavior arises from the redistribution of local stress and strain fields during matrix relaxation, which alters the electromechanical contribution of the piezoelectric phase. This case, therefore, provides a stringent test of whether the PDMN can capture the history-dependent redistribution of coupled local fields.

The prediction accuracy improves systematically with the hierarchical level $N$. For $N=4$, the PDMN captures the overall relaxation trend, but noticeable deviations remain, particularly in the $D_3$ response, with MRE values of $9.95\%$ for $D_3$ and $1.98\%$ for $\sigma_{33}$. Increasing the hierarchy to $N=6$ markedly improves the predictions, reducing these MRE values to $1.64\%$ and $0.41\%$, respectively. At $N=8$, the PDMN achieves excellent agreement with DNS, with the MRE of both responses falling below $1\%$. These results indicate that a higher hierarchical level enriches the microstructural representation capacity of the PDMN and thereby improves its predictive accuracy, particularly when extrapolating to nonlinear coupled responses.

\begin{table}[htbp]
\centering
\caption{Mean-relative and max-relative error metrics of the PDMN predictions for the stress relaxation response in Case III.}
\label{tab:case3_stress_relaxation_errors}
\renewcommand{\arraystretch}{1.15}
\setlength{\tabcolsep}{6pt}
\begin{tabular}{llccc}
\toprule
Loading path & Response & Model & MRE & MaxRE \\
\midrule
\multirow{6}{*}{Stress relaxation}
& \multirow{3}{*}{$D_3$}
& $N=8$ & $1.610460\times10^{-3}$ & $1.831650\times10^{-3}$ \\
& & $N=6$ & $1.641010\times10^{-2}$ & $1.731351\times10^{-2}$ \\
& & $N=4$ & $9.948075\times10^{-2}$ & $1.071720\times10^{-1}$ \\
\cmidrule(lr){2-5}
& \multirow{3}{*}{$\sigma_{33}$}
& $N=8$ & $7.578875\times10^{-4}$ & $8.416564\times10^{-4}$ \\
& & $N=6$ & $4.076897\times10^{-3}$ & $4.308217\times10^{-3}$ \\
& & $N=4$ & $1.977966\times10^{-2}$ & $2.132912\times10^{-2}$ \\
\bottomrule
\end{tabular}
\end{table}

\subsubsection{Computational efficiency}

The computational cost of the online prediction stage is summarized in Table~\ref{tab:computation_time_comparison}. For both phase arrangements and loading conditions, the proposed PDMN substantially reduces the computational time compared with DNS. Depending on the hierarchical level $N$ and the prescribed loading path, the achieved speed-up ranges from $1.75\times10^3$ to $4.59\times10^4$.

As expected, increasing $N$ increases the online computational cost due to the larger number of material nodes in the hierarchical network. This added cost is accompanied by improved predictive accuracy, as discussed in the preceding sections. The hierarchical level $N$ therefore provides a practical means of balancing prediction accuracy and computational efficiency: smaller networks offer faster online predictions, whereas deeper networks yield more accurate homogenized responses at increased computational cost.

Even at the largest network size considered in this study, namely $N=8$, the proposed PDMN remains more than three orders of magnitude faster than DNS for all tested cases. These results demonstrate that the proposed framework provides an efficient surrogate for nonlinear electroelastic homogenization while retaining close agreement with direct numerical simulations.

\begin{table}[htbp]
\centering
\caption{Total CPU time (in CPU-seconds) required for online predictions using the PDMN model and DNS. Speed-up is computed as the ratio $T_{\textit{DNS}} / T_{\textit{PDMN}}$.}
\label{tab:computation_time_comparison}
\renewcommand{\arraystretch}{1.2}
\setlength{\tabcolsep}{9pt}
\begin{tabular}{lcccccc}
\toprule
\multirow{2}{*}{CASE} 
& \multicolumn{2}{c}{PDMN} 
& \multicolumn{2}{c}{DNS} 
& \multicolumn{2}{c}{Speed-up} \\
\cmidrule(r){2-3} \cmidrule(r){4-5} \cmidrule(r){6-7}
& $T_{E_3}$ & $T_{\varepsilon_{33}}$ 
& $T_{E_3}$ & $T_{\varepsilon_{33}}$ 
& $E_3$-loading & $\varepsilon_{33}$-loading \\
\midrule
CASE I, N=4 & 2.7  & 4.1  & $1.22\times10^{5}$ & $1.06\times10^{5}$ & $4.52\times10^4$ & $2.59\times10^4$ \\
CASE I, N=6 & 10.2 & 12.0 & $1.22\times10^{5}$ & $1.06\times10^{5}$ & $1.20\times10^4$ & $8.83\times10^3$ \\
CASE I, N=8 & 52.5 & 55.0 & $1.22\times10^{5}$ & $1.06\times10^{5}$ & $2.32\times10^3$ & $1.93\times10^3$ \\
CASE II, N=4 & 2.6  & 2.9  & $1.16\times10^{5}$ & $1.33\times10^{5}$ & $4.46\times10^4$ & $4.59\times10^4$ \\
CASE II, N=6 & 10.9 & 11.7 & $1.16\times10^{5}$ & $1.33\times10^{5}$ & $1.06\times10^4$ & $1.14\times10^4$ \\
CASE II, N=8 & 66.4 & 54.8 & $1.16\times10^{5}$ & $1.33\times10^{5}$ & $1.75\times10^3$ & $2.43\times10^3$ \\
\bottomrule
\end{tabular}
\end{table}

\section{Conclusions} \label{sec06}

This study developed a PDMN for the nonlinear electromechanical homogenization of two-phase piezoelectric composites. To our knowledge, this is the first deep material network to embed the fully two-way electromechanical coupling of piezoelectricity, for which the generalized constitutive operator is non-symmetric and indefinite, and the mechanical and electrical interaction variables must be solved simultaneously. The network is formulated from electromechanical homogenization theory, such that its trainable parameters retain a clear microstructural interpretation and can be identified offline from linear electroelastic stiffness data. In the online stage, the coupled localization problem is reduced to solving the interaction variables associated with the internal interfaces of the network. By enforcing mechanical equilibrium and electric flux continuity through a fully coupled Newton--Raphson scheme, the homogenized stress, electric displacement, and consistent tangent operators can be efficiently recovered.

The capability of the proposed framework was examined through three representative cases. Cases~I and~II considered PVDF--LiNbO$_3$ composites with reversed phase arrangements, covering responses dominated by either the soft PVDF matrix or the nonlinear LiNbO$_3$ matrix. Case~III further assessed a viscoelastic matrix with a linear piezoelectric inclusion, demonstrating the applicability of the framework to time-dependent and history-dependent coupled responses. Across these cases, the PDMN predictions showed close agreement with DNS results for both mechanical and electric loading paths.

Future work will focus on incorporating the PDMN into two-scale simulations for device-level analysis of piezoelectric energy-harvesting systems. Another direction is to extend the framework to dielectric composites used in semiconductor packaging, where the effective electromechanical and dielectric responses of heterogeneous microstructures are closely related to device performance and reliability.

\section*{CRediT authorship contribution statement}

Ting-Ju Wei: Writing -- original draft, Software, Visualization, Methodology, Investigation, Formal analysis, Data curation, Conceptualization. Yen-Ming Lu: Software, Visualization, Investigation, Formal analysis, Data curation. Chuin-Shan Chen: Writing -- review \& editing, Validation, Supervision, Project administration, Funding acquisition, Conceptualization.

\section*{Declaration of Competing Interest}
The authors declare that they have no known competing financial interests or personal relationships that could have appeared to influence the work reported in this paper.

\section*{Acknowledgements}
This work is supported by the National Science and Technology Council, Taiwan, under Grant 111-2221-E-002-054-MY3 and 112-2221-E-007-028. We are grateful for the computational resources and support from the National Center for Research on Earthquake Engineering (NCREE), NIAR, Taiwan, NTUCE-NCREE Joint Artificial Intelligence Research Center, and the National Center of High-performance Computing (NCHC). 

\section*{Data availability}
Data will be made available on request.








% \end{thebibliography}

\bibliographystyle{unsrt}  

\bibliography{references}

@article{wei2026deep,
  title={Deep Material Network: Overview, Applications, and Current Directions},
  author={Wei, Ting-Ju and Wan, Wen-Ning and Chen, Chuin-Shan},
  journal={Multiscale Science and Engineering},
  pages={1--19},
  year={2026},
  publisher={Springer}
}

@article{srinivas2026rapid,
  title={Rapid Offline Training for Deep Material Networks via a displacement-based laminate formulation and a novel sampling technique for a compliance-based fatigue model},
  author={Srinivas, Pavan Bhat Keelanje and Kabel, Matthias and Schneider, Matti},
  journal={Computer Methods in Applied Mechanics and Engineering},
  volume={449},
  pages={118517},
  year={2026},
  publisher={Elsevier}
}

@article{wei2025foundation2,
  title={Foundation Model for Polycrystalline Material Informatics},
  author={Wei, Ting-Ju and Chen, Chuin-Shan},
  journal={arXiv preprint arXiv:2512.06770},
  year={2025}
}

@article{wei2025crystallographic,
  title={Crystallographic Texture-Generalizable Orientation-Aware Interaction-Based Deep Material Network for Polycrystal Modeling and Texture Evolution},
  author={Wei, Ting-Ju and Su, Tung-Huan and Chen, Chuin-Shan},
  journal={arXiv preprint arXiv:2512.06779},
  year={2025}
}

@article{ liu2019deep,
Author = {Liu, Zeliang and Wu, C. T. and Koishi, M.},
Title = {A deep material network for multiscale topology learning and accelerated
   nonlinear modeling of heterogeneous materials},
Journal = {Computer Methods in Applied Mechanics and Engineering},
Year = {2019},
Volume = {345},
Pages = {1138-1168},
Month = {MAR 1},
DOI = {10.1016/j.cma.2018.09.020},
ISSN = {0045-7825},
EISSN = {1879-2138},
ResearcherID-Numbers = {Liu, Zeliang/AAB-7221-2019
   Liu, Zeliang/U-8269-2018},
ORCID-Numbers = {Liu, Zeliang/0000-0002-3264-8907},
Unique-ID = {WOS:000456953900047},
}

@article{ liu2019exploring,
Author = {Liu, Zeliang and Wu, C. T.},
Title = {Exploring the 3D architectures of deep material network in data-driven
   multiscale mechanics},
Journal = {Journal of the Mechanics and Physics of Solids},
Year = {2019},
Volume = {127},
Pages = {20-46},
Month = {JUN},
DOI = {10.1016/j.jmps.2019.03.004},
ISSN = {0022-5096},
EISSN = {1873-4782},
ResearcherID-Numbers = {Liu, Zeliang/AAB-7221-2019
   Liu, Zeliang/U-8269-2018},
ORCID-Numbers = {Liu, Zeliang/0000-0002-3264-8907},
Unique-ID = {WOS:000466833900002},
}

@article{gajek2020micromechanics,
  title={On the micromechanics of deep material networks},
  author={Gajek, Sebastian and Schneider, Matti and B{\"o}hlke, Thomas},
  journal={Journal of the Mechanics and Physics of Solids},
  volume={142},
  pages={103984},
  year={2020},
  publisher={Elsevier}
}

@article{noels2022micromechanics,
title = {Micromechanics-based material networks revisited from the interaction viewpoint; robust and efficient implementation for multi-phase composites},
journal = {European Journal of Mechanics - A/Solids},
volume = {91},
pages = {104384},
year = {2022},
issn = {0997-7538},
doi = {https://doi.org/10.1016/j.euromechsol.2021.104384},
author = {Van Dung Nguyen and Ludovic Noels},
}

@article{noels2022interaction,
title = {Interaction-based material network: A general framework for (porous) microstructured materials},
journal = {Computer Methods in Applied Mechanics and Engineering},
volume = {389},
pages = {114300},
year = {2022},
issn = {0045-7825},
doi = {https://doi.org/10.1016/j.cma.2021.114300},
author = {Van Dung Nguyen and Ludovic Noels},
}

@misc{DYNA_website,
  author       = {{Computational and Multi-scale Mechanics Group}},
  title        = {Training on LS-DYNA},
  year         = {2026},
  howpublished = {\url{https://www.lstc-cmmg.org/ex-rve\#dataItem-jinh49ly}},
  note         = {Accessed: 23 April 2026}
}

@article{huang2022microstructure,
title = {Microstructure-guided deep material network for rapid nonlinear material modeling and uncertainty quantification},
journal = {Computer Methods in Applied Mechanics and Engineering},
volume = {398},
pages = {115197},
year = {2022},
issn = {0045-7825},
doi = {https://doi.org/10.1016/j.cma.2022.115197},
author = {Tianyu Huang and Zeliang Liu and C.T. Wu and Wei Chen},

}

@article{jean2024graph,
Author = {Jean, Jimmy Gaspard and Su, Tung-Huan and Huang, Szu-Jui and Wu,
   Cheng-Tang and Chen, Chuin-Shan},
Title = {Graph-enhanced deep material network: multiscale materials modeling with
   microstructural informatics},
Journal = {Computational Mechanics},
Year = {2024},
Month = {2024 MAY 18},
DOI = {10.1007/s00466-024-02493-1},
EarlyAccessDate = {MAY 2024},
ISSN = {0178-7675},
EISSN = {1432-0924},
ORCID-Numbers = {Jean, Jimmy Gaspard/0000-0001-5132-9592
   Chen, Chuin-Shan/0000-0001-5281-2535},
Unique-ID = {WOS:001226568400001},
}

@article{shin2024deep,
  title={A deep material network approach for predicting the thermomechanical response of composites},
  author={Shin, Dongil and Alberdi, Ryan and Lebensohn, Ricardo A and Dingreville, R{\'e}mi},
  journal={Composites Part B: Engineering},
  volume={272},
  pages={111177},
  year={2024},
  publisher={Elsevier}
}

@article{shin2024deep2,
  title={Deep material network for thermal conductivity problems: Application to woven composites},
  author={Shin, Dongil and Creveling, Peter Jefferson and Roberts, Scott Alan and Dingreville, R{\'e}mi},
  journal={Computer Methods in Applied Mechanics and Engineering},
  volume={431},
  pages={117279},
  year={2024},
  publisher={Elsevier}
}

@article{gajek2021fe,
title = {An FE–DMN method for the multiscale analysis of short fiber reinforced plastic components},
journal = {Computer Methods in Applied Mechanics and Engineering},
volume = {384},
pages = {113952},
year = {2021},
issn = {0045-7825},
doi = {https://doi.org/10.1016/j.cma.2021.113952},
author = {Sebastian Gajek and Matti Schneider and Thomas Böhlke},

}

@article{wei2023ls,
Author = {Wei, Haoyan and Wu, C. T. and Hu, Wei and Su, Tung-Huan and Oura,
   Hitoshi and Nishi, Masato and Naito, Tadashi and Chung, Stan and Shen,
   Leo},
Title = {LS-DYNA Machine Learning-Based Multiscale Method for Nonlinear Modeling
   of Short Fiber-Reinforced Composites},
Journal = {Journal of Engineering Mechanics},
Year = {2023},
Volume = {149},
Number = {3},
Month = {MAR 1},
DOI = {10.1061/JENMDT.EMENG-6945},
Article-Number = {04023003},
ISSN = {0733-9399},
EISSN = {1943-7889},
ResearcherID-Numbers = {Nishi, Masato/W-9540-2019
   Wei, Haoyan/J-4932-2019
   },
ORCID-Numbers = {Wei, Haoyan/0000-0002-9394-6457},
Unique-ID = {WOS:000914167000007},
}

@article{gajek2021efficient,
author = {Gajek, Sebastian and Schneider, Matti and Böhlke, Thomas},
title = {Efficient two-scale simulations of microstructured materials using deep material networks},
journal = {PAMM},
volume = {21},
number = {1},
pages = {e202100069},
doi = {https://doi.org/10.1002/pamm.202100069},

year = {2021}
}

@article{gajek2022fe,
  title={An FE-DMN method for the multiscale analysis of thermomechanical composites},
  author={Gajek, Sebastian and Schneider, Matti and B{\"o}hlke, Thomas},
  journal={Computational Mechanics},
  volume={69},
  number={5},
  pages={1087--1113},
  year={2022},
  publisher={Springer}
}

@article{li2024micromechanics,
title = {Micromechanics-informed parametric deep material network for physics behavior prediction of heterogeneous materials with a varying morphology},
journal = {Computer Methods in Applied Mechanics and Engineering},
volume = {419},
pages = {116687},
year = {2024},
issn = {0045-7825},
doi = {https://doi.org/10.1016/j.cma.2023.116687},
author = {Tianyi Li},
}

@article{wu2025stochastic,
  title={Stochastic deep material networks as efficient surrogates for stochastic homogenisation of non-linear heterogeneous materials},
  author={Wu, Ling and Noels, Ludovic},
  journal={Computer Methods in Applied Mechanics and Engineering},
  volume={441},
  pages={117994},
  year={2025},
  publisher={Elsevier}
}

@article{robertson2025microstructure,
  title={Microstructure-based Variational Neural Networks for Robust Uncertainty Quantification in Materials Digital Twins},
  author={Robertson, Andreas E and Inman, Samuel B and Lenau, Ashley T and Lebensohn, Ricardo A and Shin, Dongil and Boyce, Brad L and Dingreville, Remi M},
  journal={arXiv preprint arXiv:2512.18104},
  year={2025}
}

@article{wan2024decoding,
  title={Decoding material networks: exploring performance of deep material network and interaction-based material networks},
  author={Wan, Wen-Ning and Wei, Ting-Ju and Su, Tung-Huan and Chen, Chuin-Shan},
  journal={Journal of Mechanics},
  volume={40},
  pages={796--807},
  year={2024},
  publisher={Oxford University Press}
}

@article{su2022multiscale,
    author = {Su, Tung-Huan and Huang, Szu-Jui and Jean, Jimmy Gaspard and Chen, Chuin-Shan},
    title = {Multiscale computational solid mechanics: data and machine learning},
    journal = {Journal of Mechanics},
    volume = {38},
    pages = {568-585},
    year = {2022},
    month = {11},
    issn = {1811-8216},
    doi = {10.1093/jom/ufac037},
}

@article{wei2025foundation,
  title={Foundation model for composite microstructures: Reconstruction, stiffness, and nonlinear behavior prediction},
  author={Wei, Ting-Ju and Chen, Chuin-Shan},
  journal={Materials \& Design},
  volume={257},
  pages={114397},
  year={2025},
  publisher={Elsevier}
}

@article{sony2019literature,
  title={A literature review of next-generation smart sensing technology in structural health monitoring},
  author={Sony, Sandeep and Laventure, Shea and Sadhu, Ayan},
  journal={Structural Control and Health Monitoring},
  volume={26},
  number={3},
  pages={e2321},
  year={2019},
  publisher={Wiley Online Library}
}

@article{ferreira2022embedded,
  title={Embedded sensors for structural health monitoring: methodologies and applications review},
  author={Ferreira, Pedro M and Machado, Miguel A and Carvalho, Marta S and Vidal, Catarina},
  journal={Sensors},
  volume={22},
  number={21},
  pages={8320},
  year={2022},
  publisher={MDPI}
}

@article{mahapatra2021piezoelectric,
  title={Piezoelectric materials for energy harvesting and sensing applications: roadmap for future smart materials},
  author={Mahapatra, Susmriti Das and Mohapatra, Preetam Chandan and Aria, Adrianus Indrat and Christie, Graham and Mishra, Yogendra Kumar and Hofmann, Stephan and Thakur, Vijay Kumar},
  journal={Advanced Science},
  volume={8},
  number={17},
  pages={2100864},
  year={2021},
  publisher={Wiley Online Library}
}

@book{arnau2004piezoelectric,
  title={Piezoelectric transducers and applications},
  author={Arnau, Antonio and Soares, David},
  volume={2004},
  year={2004},
  publisher={Springer}
}

@article{adeniyi2021multi,
  title={Multi-scale finite element analysis of effective elastic property of sisal fiber-reinforced polystyrene composites},
  author={Adeniyi, Adewale George and Adeoye, Samson Akorede and Onifade, Damilola Victoria and Ighalo, Joshua O},
  journal={Mechanics of Advanced Materials and Structures},
  volume={28},
  number={12},
  pages={1245--1253},
  year={2021},
  publisher={Taylor \& Francis}
}

@article{koutsawa2018overall,
  title={Overall properties of piezoelectric composites with spring-type imperfect interfaces using the mechanics of structure genome},
  author={Koutsawa, Yao},
  journal={Composites Part B: Engineering},
  volume={153},
  pages={337--345},
  year={2018},
  publisher={Elsevier}
}

@article{abedi2020effective,
  title={An effective method for hybrid CNT/GNP dispersion and its effects on the mechanical, microstructural, thermal, and electrical properties of multifunctional cementitious composites},
  author={Abedi, Mohammadmahdi and Fangueiro, Raul and Correia, Antonio Gomes},
  journal={Journal of Nanomaterials},
  volume={2020},
  number={1},
  pages={6749150},
  year={2020},
  publisher={Wiley Online Library}
}

@article{habib2022review,
  title={A review of ceramic, polymer and composite piezoelectric materials},
  author={Habib, Mahpara and Lantgios, Iza and Hornbostel, Katherine},
  journal={Journal of Physics D: Applied Physics},
  volume={55},
  number={42},
  pages={423002},
  year={2022},
  publisher={IOP Publishing}
}

@book{heywang2008piezoelectricity,
  title={Piezoelectricity: evolution and future of a technology},
  author={Heywang, Walter and Lubitz, Karl and Wersing, Wolfram},
  volume={114},
  year={2008},
  publisher={Springer Science \& Business Media}
}

@article{yazdanparast2023determining,
  title={Determining in-plane material properties of square core cellular materials using computational homogenization technique},
  author={Yazdanparast, Reza and Rafiee, Roham},
  journal={Engineering with Computers},
  volume={39},
  number={1},
  pages={373--386},
  year={2023},
  publisher={Springer}
}

@article{torquato2021nonlocal,
  title={Nonlocal effective electromagnetic wave characteristics of composite media: Beyond the quasistatic regime},
  author={Torquato, Salvatore and Kim, Jaeuk},
  journal={Physical Review X},
  volume={11},
  number={2},
  pages={021002},
  year={2021},
  publisher={APS}
}

@article{li1998micromechanics,
  title={Micromechanics of magnetoelectroelastic composite materials: average fields and effective behavior},
  author={Li, Jiang Yu and Dunn, Martin L},
  journal={Journal of Intelligent Material Systems and Structures},
  volume={9},
  number={6},
  pages={404--416},
  year={1998},
  publisher={TECHNOMIC PUBLISHING CO., INC. 851 New Holland Ave., Box 3535, Lancaster, PA~…}
}

@article{wu2000closed,
  title={Closed-form solutions for the magnetoelectric coupling coefficients in fibrous composites with piezoelectric and piezomagnetic phases},
  author={Wu, Tsung-Lin and Huang, Jin H},
  journal={International Journal of Solids and Structures},
  volume={37},
  number={21},
  pages={2981--3009},
  year={2000},
  publisher={Elsevier}
}

@article{wang2022extended,
  title={Extended locally exact homogenization theory for effective coefficients and localized responses of piezoelectric composites},
  author={Wang, Guannan and Wang, Weijian and Liu, Cheng and Chen, Weiqiu and Mei, Yue},
  journal={Advanced Engineering Materials},
  volume={24},
  number={5},
  pages={2101194},
  year={2022},
  publisher={Wiley Online Library}
}

@article{chatzigeorgiou2019micromechanical,
  title={Micromechanical method for effective piezoelectric properties and electromechanical fields in multi-coated long fiber composites},
  author={Chatzigeorgiou, George and Javili, Ali and Meraghni, Fodil},
  journal={International Journal of Solids and Structures},
  volume={159},
  pages={21--39},
  year={2019},
  publisher={Elsevier}
}

@article{martinez2017homogenization,
  title={Homogenization of porous piezoelectric materials},
  author={Mart{\'\i}nez-Ayuso, Germ{\'a}n and Friswell, Michael I and Adhikari, Sondipon and Khodaparast, Hamed Haddad and Berger, Harald},
  journal={International Journal of solids and Structures},
  volume={113},
  pages={218--229},
  year={2017},
  publisher={Elsevier}
}

@article{wei2025orientation,
title = {Orientation-aware interaction-based deep material network in polycrystalline materials modeling},
journal = {Computer Methods in Applied Mechanics and Engineering},
volume = {441},
pages = {117977},
year = {2025},
issn = {0045-7825},
doi = {https://doi.org/10.1016/j.cma.2025.117977},
author = {Ting-Ju Wei and Tung-Huan Su and Chuin-Shan Chen},
}

@article{tassi2023nonlinear,
  title={Nonlinear micromechanical modeling of fully coupled piezo-elastic composite under large deformation and high electric field},
  author={Tassi, Nada and Azrar, Lahcen and Fakri, Nadia and Aljinaidi, Abdulmalik},
  journal={Composite Structures},
  volume={315},
  pages={116991},
  year={2023},
  publisher={Elsevier}
}

@article{noorizadegan2022piezo,
  title={A novel local radial basis function collocation method for multi-dimensional piezoelectric problems},
  author={Noorizadegan, Amir and Young, D. L. and Chen, Chuin-Shan},
  journal={Journal of Intelligent Material Systems and Structures},
  volume={33},
  number={12},
  pages={1574--1587},
  year={2022},
  publisher={SAGE Publications}
}

@article{shieh2007switching,
  title={{Switching characteristics of MPB compositions of (Bi0.5Na0.5)TiO3--BaTiO3--(Bi0.5K0.5)TiO3 lead-free ferroelectric ceramics}},
  author={Shieh, J. and Wu, K. C. and Chen, C. S.},
  journal={Acta Materialia},
  volume={55},
  pages={3081--3087},
  year={2007},
  publisher={Elsevier}
}

@article{shieh2010influence,
  title={{Influence of phase composition on electrostrains of doped (Bi0.5Na0.5)TiO3-BaTiO3-(Bi0.5K0.5)TiO3 lead-free ferroelectric ceramics}},
  author={Shieh, J. and Lin, Y. C. and Chen, C. S.},
  journal={Smart Materials and Structures},
  volume={19},
  number={9},
  pages={094007},
  year={2010},
  publisher={IOP Publishing}
}

\end{document}